\documentclass{article}
\usepackage{amssymb,amsmath,amsthm}
\usepackage{mathtools} 
\usepackage{etoolbox}%
\AtBeginEnvironment{align}{\setcounter{subeqn}{0}}
\newcounter{subeqn} \renewcommand{\thesubeqn}{\theequation\alph{subeqn}}%
\newcommand{\subeqn}{%
  \refstepcounter{subeqn}
  \tag{\thesubeqn}
}
\usepackage{xfrac} 
\usepackage{upgreek} 
\usepackage{a4wide}
\usepackage{pgf,tikz}
\usepackage{geometry}
\usepackage[utf8]{inputenc}
\usepackage[english]{babel}
\usepackage{geometry}
\usepackage{array}
\usepackage{breqn}
\usepackage{physics}
\usetikzlibrary{arrows}
\usepackage{epsfig}
\usepackage{color}

\usepackage{hyperref}

\usepackage{authblk}

\usepackage{subcaption}

\renewcommand{\theequation}{\arabic{section}.\arabic{equation}}

\newcommand{\beq}{\begin{equation}}  
\newcommand{\eeq}{\end{equation}}  
\newcommand{\bea}{\begin{eqnarray}} 
\newcommand{\eea}{\end{eqnarray}}   
\newcommand{\bear}{\begin{array}}  
\newcommand{\eear}{\end{array}} 

\newtheorem{thm}{Theorem}[section] 

\newtheorem{propn}[thm]{Proposition}

\newtheorem{lem}[thm]{Lemma}

\newenvironment{prf}{\trivlist \item [\hskip 
\labelsep {\bf Proof:}]\ignorespaces}{\qed \endtrivlist}

\theoremstyle{definition}

\newtheorem{remark}[thm]{Remark}

\newcommand{\Z}{{\mathbb Z}}
\newcommand{\C}{{\mathbb C}}
\newcommand{\R}{{\mathbb R}}

\newcommand{\rP}{\mathrm{P}}
\newcommand{\rN}{\mathrm{N}}
\newcommand{\rA}{\mathrm{A}}
\newcommand{\ra}{\mathrm{a}}
\newcommand{\rB}{\mathrm{B}}
\newcommand{\rb}{\mathrm{b}}
\newcommand{\rC}{\mathrm{C}}
\newcommand{\rD}{\mathrm{D}}

\newcommand{\rd}{\mathrm{d}}

\newcommand\la{{\lambda}}

\newcommand\ka{{\kappa}}
\newcommand\al{{\alpha}}
\newcommand\be{{\beta}}

\newcommand\om{{\omega}}

\DeclareMathOperator{\Pic}{Pic}

\newcommand\tmu{\tilde{\mu}}

\newcommand\bv{{\bf v}}
\newcommand\bx{{\bf x}}
\newcommand\by{{\bf y}}
\newcommand\bd{{\bf d}}
\newcommand\bu{{\bf u}}

\newcommand\tx{\tilde{x}}
\newcommand\im{\mathrm{im}\,} 
\newcommand\hphi{\hat{\varphi}}
\newcommand\hatom{\hat{\omega}}

\newcommand\boa{{\bf a}}
\newcommand\bob{{\bf b}}
\newcommand\boc{{\bf c}}
\newcommand\boe{{\bf e}}

\newcommand\bV{{\bf V}}
\newcommand\bZ{{\bf Z}}
\newcommand\bW{{\bf W}}

\title{Commuting integrable maps from a deformed D$_4$ cluster algebra}
\author[1]{Andrew N. W. Hone\footnote{Corresponding author e-mail: A.N.W.Hone@kent.ac.uk}} 
\author[2]{Wookyung Kim}
\author[2]{Takafumi Mase}

\affil[1]{School of Engineering, Mathematics 
\&  Physics \protect\\ 
University of Kent,
Canterbury CT2 7FS, U.K.
}

\affil[2]{Graduate School of Mathematical Sciences  \protect\\ 
University of Tokyo, 3-8-1 Komaba, Tokyo 153-8914, Japan.
}
\date{\today}

\begin{document}

\maketitle

\begin{abstract} 
A cluster algebra is a special type of commutative algebra whose generators are defined 
iteratively from an initial seed via a process called mutation. The simplest cluster algebras are those of finite type, whose seeds are classified (up to mutations) by 
Dynkin diagrams of simple Lie algebras. Certain combinations of cluster mutations produce discrete dynamical systems (cluster maps), which in finite type 
have the property known as Zamolodchikov periodicity, meaning that every orbit is periodic with the same period. 

The basic combinatorial data for a cluster algebra is provided by an exchange matrix, which defines an associated log-canonical closed 2-form (presymplectic form).  
In work by one of us with Kouloukas, we showed that it is possible to deform cluster mutations 
in such a way that the underlying presymplectic form remains invariant under the cluster map. Moreover, for finite type algebras of low rank, we showed that there are parametrized families of deformations which, while destroying the Zamolodchikov periodicity of the 
original maps, preserve the feature of being Liouville integrable. 
Furthermore, these deformed integrable maps can be lifted to an enlarged phase space with a cluster structure, on which the maps are 
given by compositions of standard cluster mutations.  

In this paper we revisit an integrable map of the plane
which we obtained recently as a two-parameter family of 
deformed mutations in the cluster algebra of type D$_4$. The rational first integral for this map defines an invariant foliation of the plane by level curves, and we explain how this corresponds to a rational elliptic surface of rank 2. 
This leads us to construct another (independent) integrable map, commuting with the first, such that both maps lift to compositions of 
mutations in an enlarged cluster algebra,  whose  underlying quiver is equivalent to the one found 
by Okubo for the $q$-Painlev\'e VI equation. The degree growth of the two commuting maps is calculated in two different ways: firstly, from the tropical (max-plus) equations for the d-vectors of the cluster variables; 
and secondly, by constructing the minimal space of initial conditions for the two maps, via blowing up $\mathbb{P}^1 \times \mathbb{P}^1$. 
\end{abstract}

\section{Introduction}

\setcounter{equation}{0}

There are two main areas that served as sources of inspiration and motivation for Fomin and Zelevinsky's development of cluster algebras: the first was Lie theory, specifically Lusztig's conjectures on the canonical basis; and the second was discrete dynamical systems, in particular Somos sequences and certain related nonlinear recurrences 
exhibiting the Laurent phenomenon. However, there is a third area of inspiration that incorporated elements of both Lie theory and discrete dynamics, namely the Zamolodchikov conjecture on the periodicity of solutions of Y-systems, which are special systems of difference equations arising from the thermodynamic Bethe ansatz for ADE scattering theories. In the original setting 
of \cite{zam},  Zamolodchkov's observation was that,  in the dynamics of the Y-system associated with a simply-laced Dynkin diagram of finite type, every orbit is periodic with the same period, namely the Coxeter number plus two. 
An analogous type of periodicity is observed in T-systems \cite{kun}, which appear in the closely related context of Yangians and quantum affine algebras.
Within the framework of cluster algebras, Y-systems arise as certain compositions of coefficient mutations, while T-systems come from cluster mutations. The latter will be our focus in this paper, so here, for the sake of convenience, we work with the coefficient-free case. 

A coefficient-free cluster algebra ${\cal A} ( \bx, B)$ of rank $N$ is constructed from a seed  $( \bx, B)$, which consists of the $N$-tuple $\bx = (x_1,\ldots,x_N)$ of cluster variables, together with 
the exchange matrix $B=(b_{ij})\in \operatorname{Mat}_N(\Z)$. 
In general, $B$ is just required to be skew-symmetrizable, but for our current purposes it is sufficient to consider only skew-symmetric $B$, in which case $B$ is equivalent to a quiver $Q$ 
without self loops and 2-cycles, such that  
each entry $b_{ij}$ counts the number of arrows 
$i\to j$ minus the number of arrows 
$j\to i$ between the vertices of $Q$, labelled with the indices $1,\ldots,N$. For each choice of vertex $k$, the matrix mutation $\mu_k$ is the transformation of $B$ that produces a new exchange matrix $B'=\mu_k(B)=(b_{ij}')$, whose entries are defined by 
\begin{equation}\label{matrixmu}
        b^{'}_{ij} =
        \begin{cases}
           - b_{ij} & \text{if} \ i= k \ \text{or} \ j=k  \\ 
            b_{ij} + \frac{1}{2} \qty(\abs{b_{ik}}b_{kj} + b_{ik}\abs{b_{kj}})& \text{otherwise}  .\\ 
        \end{cases}
    \end{equation}
(For brevity, we omit the corresponding description of this in terms of mutation of the quiver $Q$.) Furthermore, the associated cluster mutation produces a new cluster $\bx' = \mu_k (\bx) =(x_j')$, whose components for all indices except $k$ are the same, so that $x_j'=x_j$ for $j\neq k$, while for index $k$ the new variable 
$x_k'$ is a rational expression in the cluster variables from the initial seed, being given by the 
exchange relation 
\beq \label{exchange}
x_k'x_k = \prod_{j=1}^N x_j^{[b_{jk}]_+}
+ \prod_{j=1}^N x_j^{[-b_{jk}]_+} 
\eeq 
(where for $b\in\R$ we set $[b]_+=\max (b,0)$). 
Overall, the mutation $\mu_k$ generates a new seed $(\bx',B')=\mu_k (\bx ,B)$. Then the cluster algebra ${\cal A} ( \bx, B)$ is defined to be 
the $\Z$-algebra generated by all cluster variables from all possible mutations applied to the initial seed $( \bx, B)$. 

For almost all choices of exchange matrix $B$ in 
the initial seed, the cluster algebra 
${\cal A} ( \bx, B)$ has infinitely many generators. 
The exceptions to this rule are the cluster algebras of finite type, classified in 
\cite{fz2}, which (up to mutation equivalence) 
are given by exchange matrices $B$ that are built from finite type Dynkin quivers. 
These finite Dynkin type algebras also  
lie behind Zamolodchikov periodicity \cite{fzYsys}. For the associated 
discrete dynamical systems that are defined by 
compositions of cluster mutations (T-systems), a key feature to be used in what follows is the 
log-canonical 
presymplectic form associated 
with the seed $(\bx,B)$, that is 
\beq\label{pres} 
\om = \sum_{1\leq i< j\leq N} \frac{b_{ij}}{x_ix_j} \,\rd x_i \wedge \rd x_j , 
\eeq 
which transforms covariantly under mutations, so that under the action of the mutation 
$\mu_k $ it is transformed to 
$$
\om' =\mu_k^*\,\om = 
\sum_{1\leq i< j\leq N} \frac{b_{ij}'}{x_i'x_j'} \,\rd x_i' \wedge \rd x_j' . 
$$
In the finite type case, Zamolodchikov periodicity entails that there is a sequence of mutations whose composition leaves the matrix $B$ invariant, as well as  returning the original seed $\bx$ to itself; hence,  overall such a transformation also preserves the closed 2-form \eqref{pres}. 

Our point of departure for the rest of the article is the composition 
$\mu_4\mu_3\mu_2\mu_1$  
of four mutations in the cluster algebra of type 
D$_4$, 
given by the following sequence of exchange relations: 
\beq \label{D4orig}\begin{array}{rcl}
\mu_{1} : (x_1,x_2,x_3,x_4) \mapsto (x_1',x_2,x_3,x_4), \qquad x_1 x_1' &=& 1 + x_2 \\
\mu_{2} :(x_1',x_2,x_3,x_4) \mapsto (x_{1}',x_{2}',x_3,x_4), \qquad x_2 x_2' &=& 1 +  x_3 x_4 x_{1}'\\
\mu_{3} : (x_{1}',x_{2}',x_3,x_4) \mapsto (x_{1}',x_{2}',x_{3}',x_4), \qquad  x_3 x_{3}'& = & 1 +  x_2'\\
\mu_{4} : (x_1',x_2',x_3',x_4) \mapsto (x_{1}',x_{2}',x_{3}',x_{4}'), \qquad x_4 x_4' &= & 1 +  x_{2}' 
.
\end{array}
\eeq 
Overall this defines a 4D birational map 
$\varphi=\mu_4\mu_3\mu_2\mu_1$, that is 
\beq \label{D4map}
\varphi: \, 
(x_1,x_2,x_3,x_4) \mapsto 
(x_1',x_2',x_3',x_4'), 
\eeq 
of which every orbit has period 
$4=\frac{1}{2} (6+2)$ (half the 
Zamolodchikov period). 
Viewing it as an automorphism of the field of fractions 
$\C (x_1,x_2,x_3,x_4)$, the components of each of the iterates of the map are cluster variables, 
so by the Laurent phenomenon each of these components belongs to the ring $\cal R$ 
of Laurent 
polynomials in the seed variables, with 
integer coefficients, that is 
\beq \label{lpoly}
{\cal R} = \Z [x_1^{\pm 1}, x_2^{\pm 1},x_3^{\pm 1},x_4^{\pm 1}] 
\eeq 
In particular, this means that choosing all four values of the 
initial seed variables $x_j=1$ produces an orbit 
of \eqref{D4map} consisting entirely of integers, namely 
\beq\label{intorbit}
(1,1,1,1)\mapsto 
(2,3,4,4)\mapsto 
(2,11,3,3)\mapsto 
(6,5,2,2)\mapsto 
(1,1,1,1), 
\eeq 
providing an example of a type D generalization of Coxeter's frieze patterns \cite{MGfriezes}. 

\subsection{Outline of the paper} 

In the next section, we review some results from 
\cite{hkm24} concerning deformations of the 
D$_4$ cluster map 
\eqref{D4map}, which destroy the periodicity and 
the Laurent property, but retain the feature of integrability.  The focus of the rest of the article is on a two-parameter family of deformations of \eqref{D4map} that reduces to an integrable symplectic map in the plane, 
denoted $\hat\varphi$. 
The 
latter map  
has a rational invariant that defines a pencil of curves of bidegree $(3,2)$ 
in $\mathbb{P}^1 \times \mathbb{P}^1$, and 
we show that the resulting elliptic surface has rank 2.  Since the map $\hat\varphi$ corresponds to a translation by a generator of the associated 
Mordell-Weil group, this inspired us to search for a commuting map $\hat\chi$, corresponding to  translation by a second independent generator. 

The symplectic map $\hat\varphi$ also admits a cluster structure,  lifting it to a 
map $\psi$ in 8 dimensions, where the Laurent property is restored. 
More precisely, the lifted map $\psi$ corresponds to a composition of mutations and a permutation in a cluster algebra of rank 8, with the parameters being an extra 2 frozen variables. 
In section 3, we show how the new integrable map $\hat\chi$ 
that commutes with $\hat\varphi$ 
is constructed from another  
sequence of mutations composed with a permutation in the rank 8 cluster algebra. From the analysis of the tropical versions of the exchange relations, we calculate the degree growth of cluster variables obtained via iteration of the two maps. Finally, in section 4,  we construct the common space of initial conditions for the maps $\hat\varphi$ and $\hat\chi$, by blowing up the pencil of curves at its 9 base points, before blowing down once to get a minimal surface ${\cal X}_0$. The induced linear action on the Picard lattice provides an another independent check on the degree growth of the two maps.  
The paper ends with some conclusions.

\section{Deformation of the $\rD_{4}$ cluster  map}
\setcounter{equation}{0}

The cluster algebra of type $\rD_4$ is defined by the exchange matrix 
\begin{equation}\label{D4exch}
  B= \mqty(0 & 1 & 0 & 0 \\-1 & 0 & 1 & 1 \\ 0 & -1 & 0 & 0 \\ 0 & -1 & 0 & 0) , 
\end{equation}
whose Cartan counterpart $A=A(B)$ (in the sense of \cite{fz2}) is 
the Cartan matrix for the $\rD_4$ root system, 
that is 
$$
    \mqty(2 & -1 & 0 & 0 \\-1 & 2 & -1 & -1 \\ 0 & -1 & 2 & 0 \\ 0 & -1 & 0 & 2) .
$$ 
The presymplectic form associated with (\ref{D4exch}) is 
given by 
\beq\label{d4om} 
\om =  \frac{1}{x_{1}x_{2}} \dd  x_{1} \wedge \dd x_{2}  + \frac{1}{x_{2}x_{3}} \dd  x_{2} \wedge \dd x_{3} + \frac{1}{x_{2}x_{4}} \dd x_{2} \wedge \dd  x_{4}, 
\eeq 
and this transforms covariantly under the action of mutations.  

The $\rD_4$ cluster map \eqref{D4map} is composed of the sequence of mutations \eqref{D4orig}. Since this returns the original exchange matrix \eqref{D4exch} to itself, i.e. 
$$ 
\mu_4\mu_3\mu_2\mu_1 (B) = B, 
$$
it follows that the 2-form \eqref{d4om} is invariant under the cluster map, or in other words 
$$
\varphi^* \, \om =\om.
$$
However, $\varphi$  is not quite a symplectic map, because the 2-form $\om$ is degenerate. 

The degenerate exchange matrix 
\eqref{D4exch} 
has rank 2, which means that 
one can reduce the birational map $\varphi$ from 4D to a 2D  symplectic map. 
A convenient choice of $\Z$-basis for the null space and for the image of $B$ is given by 
\begin{equation} \label{kerim}
    \ker(B) = <(1,0,0,1)^{T}, (1,0,1,0)^T>, \qquad \im(B)=<(0,1,0,0)^{T}, (-1,0,1,1)^{T}>.  
\end{equation}
It follows that the null distribution of the presymplectic form $\omega$ is spanned by the two commuting vector fields $\bv_{1} = x_{1}\partial_{x_{1}} + x_{4}\partial_{x_{4}}$ and $\bv_{2} = x_{1} \partial_{x_{1}} + x_{3} \partial_{x_{3}}$. The space of leaves of the null foliation has local coordinates 
\begin{equation}\label{leaves}
    y_1 = x_2,\quad y_2 = \frac{x_3x_4}{x_1} , 
\end{equation}
which are a pair of Laurent monomials in the initial cluster variables, whose exponents are  given 
by the basis vectors of $\im(B)$
in \eqref{kerim}. %
Then the  rational map defined by 
\beq\label{pimapD4} 
 \begin{array}{lcrcl}
    \pi&:&\C^4&\to&\C^2\\
    & &\bx=(x_1,x_2,x_3,x_4)&\mapsto&\by=(y_1,y_2) \; 
  \end{array}
\eeq 
reduces the cluster map $\varphi$ to a 
symplectic map defined on the plane, namely 
\beq\label{phihatmapD4} 
 \begin{array}{lcrcl}
    \hat{\varphi}&:&\C^2&\to&\C^2\\
    & &\by=(y_1,y_2)&\mapsto&\left(\dfrac{y_1y_2 +y_2+ 1}{y_1}, \dfrac{(y_1+1)(y_2+1)^2}{y_1^2y_2}\right) \;, 
  \end{array}
\eeq 
which is intertwined with $\varphi$ via $\pi$, so that 
\beq \label{intertw}
\hat{\varphi}\cdot\pi = \pi\cdot\varphi, \qquad \hphi^*(\hatom)=\hatom, 
\eeq 
where $\pi^*(\hatom)=\om$ is the pullback of the symplectic form 
\beq\label{symD4iy}
\hatom=\frac{1}{y_1y_2}\rd y_1\wedge \rd y_2 
\eeq 
under $\pi$. 

\begin{remark}\label{sympex}
Under the map \eqref{pimapD4}, the orbit \eqref{intorbit} projects down to the period 4 orbit 
$$
(1,1) \mapsto (3,8) \mapsto  (11,\frac{9}{2}) 
\mapsto  (5,\frac{2}{3}) \mapsto (1,1) 
$$
of the map \eqref{phihatmapD4}. 
\end{remark}

As illustrated by the preceding remark, the Zamolodchikov periodicity of the map $\varphi$ 
is inherited by the symplectic map \eqref{phihatmapD4}, which also has every orbit being period 4 ($\hat{\varphi}^4 =\mathrm{id}$), and this means 
that it is trivially integrable: in 2D, Liouville integrability requires the existence of one invariant function, but in this case any function averaged over a periodic orbit is invariant (and  two functionally independent rational invariants 
can be obtained in this way). For our purposes, the particular function of interest is 
\begin{equation}\label{K1}
    K = \sum_{i=0}^{3} (\hat{\varphi}^*)^{i}(y_1) = \frac{(1 + y_1)^3 + (2 + 5y_1 + y_1^3) y_2 + (1 + y_1)^2 y_2^2}{y_1^2y_2}, 
\end{equation}
which satisfies $\varphi^*( K) =K$ (by periodicity), i.e.\ it is invariant. 
The generic level curves $K=\,$const in the plane have genus one, and we seek to deform the cluster map while retaining this geometric feature. 

We now consider deformations of the mutations
\eqref{D4orig} in the $\rD_4$ cluster algebra, 
with parameters $\ra_j, \rb_j$ for $1\leq j\leq 4$,  taking the form 
\beq \label{D4dmaps}\begin{array}{rcl}
\mu_{1} : (x_1,x_2,x_3,x_4) \mapsto (x_1',x_2,x_3,x_4), \qquad x_1 x_1' &=& \rb_1 + \ra_1x_2 \\
\mu_{2} :(x_1',x_2,x_3,x_4) \mapsto (x_{1}',x_{2}',x_3,x_4), \qquad x_2 x_2' &=& \rb_2 + \ra_2 x_3 x_4 x_{1}'\\
\mu_{3} : (x_{1}',x_{2}',x_3,x_4) \mapsto (x_{1}',x_{2}',x_{3}',x_4), \qquad  x_3 x_{3}'& = & \rb_3 + \ra_3 x_2'\\
\mu_{4} : (x_1',x_2',x_3',x_4) \mapsto (x_{1}',x_{2}',x_{3}',x_{4}'), \qquad x_4 x_4' &= & \rb_{4} + \ra_4 x_{2}' 
.
\end{array}
\eeq 
The same letter $\varphi$ is used to denote 
the deformed cluster map given by 
the composition of the above mutations with 
parameters, that is  $\varphi= \mu_{4}\mu_{3}\mu_{2}\mu_{1}$, 
which 
reduces to the original cluster map  
(\ref{D4map}) when we fix the parameters $\ra_i=1=\rb_i$ for all $i$. 
It is convenient to reduce the number of parameters in the problem  
by rescaling each of the cluster variables independently, $x_{i} \mapsto \lambda_{i} x_{i}$, and choose the multipliers $\la_i$ so that the parameters $\ra_j$ are 
rescaled to 1,  and then the deformed map 
$\varphi$ takes the same form as 
\eqref{D4map} but with the new variables 
$x_j'$ being specified by 
%
\beq\label{D4recs} 
\begin{array}{rcl}
x_{1}' x_{1} &=& x_{2} + \rb_1 , \\   
x_{2}' x_{2} &=& x_{3} x_{4} x_{1}' + \rb_2, \\  
x_{3}' x_{3}& = &  x_{2}' + \rb_3 , \\ 
x_{4}' x_{4} &= &  x_{2}' + \rb_4
.
\end{array} 
\eeq 

For generic values of $\rb_i$,  the 4-parameter family of deformed maps does not have periodic orbits, and each component of the iterates belongs to $\C(x_1,x_2,x_3,x_4)$ but is not a Laurent polynomial: so both Zamolodchikov periodicity and the Laurent property are 
destroyed by the deformation. However, by 
the result of Theorem 1.3 in \cite{hk}, 
each member of the family preserves the same 
presymplectic form (\ref{d4om}) associated 
with the $\rD_4$ exchange matrix \eqref{D4exch}, 
i.e. $\varphi^* \, \om =\om$ still holds for all choices of the parameters $\rb_i$. 
Furthermore, the same transformation \eqref{pimapD4} 
reduces \eqref{D4recs} to a symplectic map 
in the plane, given by 
\beq\label{newphihatmapD4} 
 \begin{array}{lcrcl}
    \hat{\varphi}&:&\C^2&\to&\C^2\\
    & &\by=(y_1,y_2)&\mapsto&\left(\dfrac{(\rb_1 + y_1)y_2 + \rb_2}{y_1}, \dfrac{\big((\rb_4 + y_2)y_1 + \rb_1y_2 + \rb_2\big)\big((\rb_3 + y_2)y_1 + \rb_1y_2 + \rb_2\big)}{y_2y_1^2(\rb_1 + y_1)}\right) \;, 
  \end{array}
\eeq 
which again satisfies \eqref{intertw}, 
with the same symplectic form 
\eqref{symD4iy}. 
 
The deformed symplectic map 
does not appear to be integrable for arbitrary choices of the parameters $\rb_i$. 
However, the approach proposed in \cite{hk} is to seek a first integral $\tilde{K}$ that is a linear combination of the same monomials in $y_1,y_2$ that appear in the undeformed expression \eqref{K1}. 
Therefore,  
we suppose that there is an analogous 
invariant function 
for 
\eqref{newphihatmapD4}, taking the form 
\begin{equation}\label{firstintD4}
    \tilde{K} = y_1 + c_1 y_2 + \frac{c_2 y_1}{y_2} + \frac{c_3 y_2}{y_1} + \frac{c_4}{y_2} + \frac{c_5}{y_1} + \frac{c_6 y_2}{y_1^2} + \frac{c_7}{y_2y_1} + \frac{c_8}{y_1^2} + \frac{c_9}{y_2y_1^2}
\end{equation}
where $c_{j}$ are  parameters to be determined. Then imposing the requirement that  
$\hat{\varphi}^{*}(\tilde{K}) = \tilde{K} $ 
leads us to necessary and sufficient conditions for the map 
$\hat{\varphi}$ to be integrable with an invariant of this form, as follows. 
\begin{thm}\label{defd4}\cite{hkm24}  
For the deformed symplectic map $\hat{\varphi}$ to admit the first integral \eqref{firstintD4}, it is necessary and sufficient for the parameters $\rb_{i}$ to satisfy one  of the following sets of conditions: 
\begin{subequations}
\begin{equation}\label{eq:4}
     (1) \  \rb_2 = \rb_4 = 
\rb_1 \rb_3 ; 
\end{equation}
\begin{equation}\label{eq:5}
    (2) \  \rb_1 = \rb_2 = 
\rb_3 \rb_4 ; 
\end{equation}
\begin{equation}\label{eq:6}
    (3) \ \rb_2=\rb_3 =  
\rb_1\rb_4 .
\end{equation}
  \end{subequations}
In each of these cases, subject to the given 
conditions on the parameters $\rb_i$, 
the deformed map $\hat{\varphi}$ given by \eqref{newphihatmapD4} 
 is Liouville integrable, preserving 
the function 
\small
\begin{equation}\label{firstintD4v1}
    \tilde{K} = y_1 + y_2 + \frac{y_1}{y_2} + \frac{(\rb_1 + 1)y_2}{y_1} + \frac{\rb_3 + \rb_4 + 1}{y_2} + \frac{\rb_1 + \rb_2 + \rb_3 + \rb_4 + 1}{y_1} + \frac{\rb_1y_2}{y_1^2} + \frac{\rb_3\rb_4 + \rb_3 + \rb_4}{y_1y_2} + \frac{2\rb_2}{y_1^2} + \frac{\rb_3\rb_4}{y_1^2y_2}.
\end{equation}
\normalsize
\end{thm}

Note that the deformed mutations  (\ref{D4recs}) remain invariant under switching $x_3\leftrightarrow x_4$, $\rb_3\leftrightarrow \rb_4$, 
and similarly for the form of the reduced map  $\hat{\varphi}$ in \eqref{newphihatmapD4} and the first integral \eqref{firstintD4v1} when these last two 
parameters are switched. Hence 
$(1)$ and $(2)$ are really the only two distinct cases to consider in Theorem~\ref{defd4}, 
since (1) and (3) are equivalent to one another.
For the rest of this paper, we will focus on case (2) only. 

\subsection{Deformed $\rD_4$ symplectic map 
$\hat\varphi$ in case (2) } 

In what follows, it will be convenient to replace the coordinates and parameters in case (2) above as follows: 
$$
(y_1,y_2) \rightarrow (z,w), \qquad 
(\rb_1,\rb_2,\rb_3,\rb_4) \rightarrow 
(\al\be, \al\be, \al, \be). 
$$
With this notation, 
the particular form of the map \eqref{newphihatmapD4} that is 
specified by case (2) of 
Theorem~\ref{defd4} becomes 
\begin{equation}\label{D42map}
    \hat{\varphi}: \quad  \mqty(z \\ w) \mapsto 
\qty(\frac{(\al\be+z)w + \al\be}{z}, \quad  \frac{\big((\be + w)z + \al\be(w + 1)\big)
\big((\al + w)z + 
\al\be (w+1)\big)}{z^2w(\al\be+z)}), 
\end{equation} 
and the invariant symplectic form is given by 
\beq\label{zwom}
\hat{\om} = \frac{1}{zw}\, \rd z\wedge \rd w. 
\eeq 
Also, dropping the tilde, the first integral 
\eqref{firstintD4v1} in this case becomes 
\begin{equation}\label{D42firstint}
     {K} = z + w + \frac{z}{w} + \frac{(\al\be + 1)w}{z} + \frac{\al + \be + 1}{w} 
+ \frac{2\al\be + \al + \be + 1}{z} + \frac{\al\be w}{z^2} + \frac{\al\be + \al + \be}{zw} + \frac{2\al\be}{z^2} + \frac{\al\be}{z^2w}. 
\end{equation}
Henceforth, our main goal will be to understand the detailed structure of the integrable map 
\eqref{D42map}. 

A fixed level set $K=\ka$ of \eqref{D42firstint} 
is defined by the polynomial equation 
\small
\beq\label{Klevel}
z^3w+z^2w^2+z^3+(\al\be +1)zw^2+(\al+\be+1)z^2
+(2\al\be+\al+\be+1)zw+\al\be w^2+
(\al\be+\al+\be)z+2\al\be w+\al\be 
=\ka z^2 w.
\eeq 
\normalsize
Allowing the coordinates $z,w$ to take values 
in $\C\cup \{ \infty \} \cong\mathbb{P}^1$, the 
equation \eqref{Klevel} defines a curve of bidegree $(3,2)$ in $\mathbb{P}^1\times \mathbb{P}^1$, which has genus 1 for generic  
$\ka\in\mathbb{P}^1$. Moreover, 
a generic point $(z,w)$ in the plane specifies a unique value of $K$, so the function 
\eqref{D42firstint} defines an elliptic fibration over $\mathbb{P}^1$. 

QRT maps provide archetypal examples of integrable maps in the plane \cite{Duistermaat2010DiscreteIS}. 
However, such maps are constructed from pencils of curves of bidegree $(2,2)$ (biquadratic curves), so the map \eqref{D42map} is not of QRT type. Nevertheless, in section 4 we shall show a relation with a QRT map. 


\subsection{Rational elliptic surface}

The pencil of curves \eqref{Klevel}
defines a rational elliptic surface. Equivalently, by making a birational transformation to Weierstrass normal form, this can be viewed as an elliptic curve $E$ over 
$\C(\ka)$. The Weierstrass form is convenient for 
computing the j-invariant of this curve, which 
(with the aid of computer algebra) is found to be given by 
\beq\label{jinv}
j(E) = \frac{p_4(\ka)^3}{ (\ka+\al+\be+2)^3 (\ka+\al\be+3)^2 \, p_3(\ka)}, 
\eeq 
where 
$$ 
p_3 =\al\be\, \ka^3+\cdots,  \qquad  p_4 = \ka^4 +\cdots 
$$
are polynomials in $\ka$ of degrees 3, 4 respectively, which both have only simple roots (for generic values of $\al,\be$). 

The poles of the j-invariant correspond to the singular fibres, where the curve becomes degenerate. The 3 roots of $p_3$ give values of $\ka$ where   \eqref{Klevel} is an irreducible rational curve. The first factor in the denominator of the j-invariant gives a triple pole at $\ka=-\al-\be-2$, meaning that the 
singular fibre has multiplicity 3. This can be seen be rewriting  \eqref{Klevel} as 
$$ 
\rA(z) \,w^2 +\rB(z)\, w+\rC(z)=0, 
\quad \mathrm{with}\quad 
\rA=(z+1)(z+\al\be), \, 
\rB=\rA+\rC-(\ka+\al+\be+2)z^2, \, 
\rC=(z+1)(z+\al)(z+\be). 
$$
Then it can be verified that 
when $\ka+\al+\be+2=0$, the equation for the curve decomposes into three factors as 
$$ 
(z+1)(w+1)\, \ell_1 (z,w)=0, 
$$
where the third factor $\ell_1$ is linear in $w$; so each factor above defines a rational curve. 
Similarly, the second 
factor in the denominator of the j-invariant gives a double pole at $\ka=-\al\be-3$, and for this value of $\ka$ the equation of the curve 
decomposes as 
$$ 
(w+z+1)\, \ell_2(z,w)=0, 
$$
where the factor $\ell_2$ is also linear in $w$, so this singular fibre has multiplicity 2. Furthermore, the difference of degrees between the numerator and denominator of the j-inavriant is $12-8=4$, so there is a singular fibre of multiplicity 4 at $\ka=\infty$. It is clear that the two lines $z=0$ and $w=0$ lie in this fibre, and by considering other charts with coordinates 
$Z=1/z$ and $W=1/w$ it becomes apparent that an additional two lines, namely  $z=\infty$ and $w=\infty$, are also included, which completes the count.  

The Mordell-Weil group of the curve $E\big(\C(\ka)\big)$ 
corresponding to the pencil 
 \eqref{Klevel}
 is the set  of $\C(\ka)$-rational points, which is a finitely generated abelian group. The structure of the singular fibres allows us to determine the rank. 

\begin{propn}\label{rankX}
The Mordell-Weil group of the rational elliptic surface defined by the pencil of curves 
\eqref{Klevel} has rank 2.
\end{propn}
\begin{prf}
As noted above, there are three reducible 
singular fibres with multiplicities $2,3,4$ respectively. By Theorem 3.1 in \cite{os}, the rank of the root lattice
$T$ associated with these reducible fibres is given in terms of their multiplicities as  
$\mathrm{rk}\,T = (2-1)+(3-1)+(4-1)=6$, 
and then the rank of the elliptic surface is 
$$ 
\mathrm{rk}\, 
E\big(\C(\ka)\big) = 8 - \mathrm{rk}\,T = 2.
$$
\end{prf} 

The Mordell-Weil group can also be considered as the set of sections of the elliptic fibration. There is a zero section $s_0$, which can be regarded as a choice of identity element ${O}$ in each 
(generic) fibre. Moreover, as the birational map 
$\hat\varphi$ preserves each of the level curves 
defined by \eqref{Klevel}, it induces an infinite order automorphism on each fibre, 
which acts as addition of a point $P_1$ in the associated group law, and hence corresponds to another section, $s_1$ say. Now, because the group has rank 2, there is another independent section, $s_2$ say, corresponding to addition of a point $P_2$ in the group law, such that 
$nP_1+mP_2\neq O$ for all integers $n,m$ not both zero. Then in turn, there should be another birational map 
$\hat{\chi}$ which acts as translation by $P_2$ on each (generic) fibre, 
with $\hat{\varphi}^n \neq \hat{\chi}^m$ for all $(m,n)\neq (0,0)$. 
We do not know of any general algorithm to 
find such a  map $\hat{\chi}$, but in the next section we will succeed in constructing one in terms of the cluster structure on an enlarged phase space. 

\section{Cluster structure and commuting map}
\setcounter{equation}{0}

As noted previously, while the original 
$\rD_4$ cluster map generates Laurent polynomials 
in the ring \eqref{lpoly}, the deformation parameters in \eqref{D4recs} destroy the Laurent property. However, in many cases it is possible to perform ``Laurentification'', meaning that there is a lift to an enlarged phase space, with new coordinates in which the Laurent property holds. We recall how to do this for 
the deformed $\rD_4$ symplectic map 
$\hat\varphi$ by presenting an associated 
extended cluster structure, which is the content of Theorem 5.6 in \cite{hkm24},  and then proceed 
to use the same extended cluster variables to 
construct a new integrable map $\hat\chi$ which commutes with $\hat\varphi$ . 

\subsection{Extended cluster algebra 
associated with the map $\hat\varphi$}
In \cite{hkm24} we used the singularity structure of the symplectic map (\ref{D42map}) 
to derive a lift to a cluster algebra of rank 8, 
with initial cluster 
$\tilde{\bx}=(\tx_1,\tx_2,\ldots,\tx_8)$, defined by a particular rational map $\tilde{\pi}:\C^8\rightarrow \C^2$, that is 
\beq\label{pitdef} 
 \tilde{\pi}: \qquad z =\frac{\tx_3}{\tx_{5}\tx_6}, \qquad 
w = \frac{\tx_2\tx_4\tx_8}{\tx_1\tx_6\tx_{7}}. 
\eeq 
After pulling back the symplectic form \eqref{zwom} by this map, we obtain the degenerate 2-form 
\beq\label{tom}
    \tilde{\omega} = \tilde{\pi}^{*}( \hat{\omega}) = \sum_{1\leq i<j\leq 8} \frac{\tilde{b}_{ij}}{\tilde{x}_{i}\tilde{x}_{j}} \dd \tilde{x}_{i} \wedge \dd \tilde{x}_{j} . 
\eeq 
Then the skew-symmetric matrix of coefficients $\tilde{B}=(\tilde{b}_{ij})$ defined by the 
 presymplectic form  $\tilde{\omega}$, as above, 
 defines the exchange matrix in a seed 
$(\tilde{\bx},\tilde{B}) $ of this algebra. 

\begin{figure}[h]
\centering
\epsfig{file={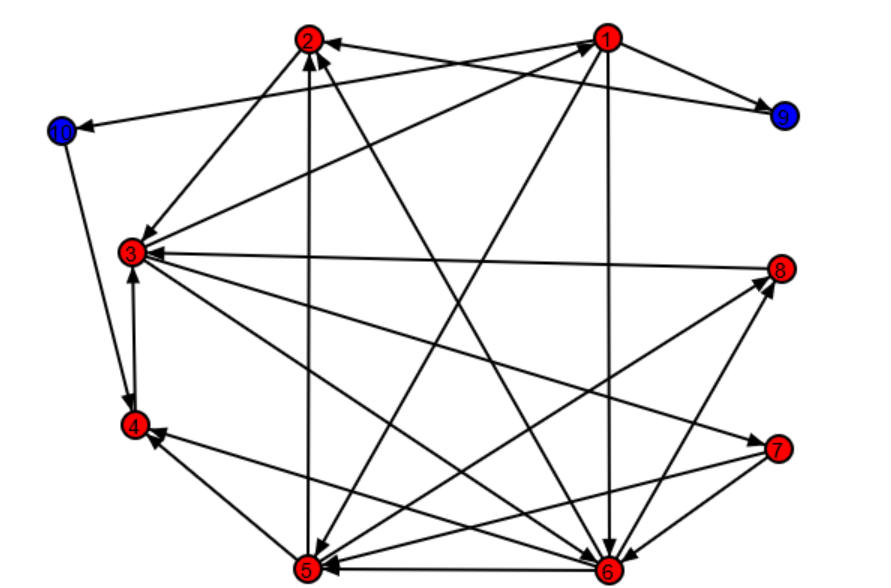}, height=3in, width=4.8in}
\caption{Extended quiver associated with the cluster map $\psi$}
\label{quiverD4ii}
\end{figure}

However, because of the parameters appearing 
in the deformation, we cannot consider the coefficient-free cluster algebra 
${\cal A}(\tilde{\bx},\tilde{B}) $ alone, 
but rather we must consider an enlarged 
cluster algebra of geometric type, with 
an extended seed 
$$ 
\hat{\bx} = (\tx_1\,\tx_2,\ldots, \tx_8,\tx_9, \tx_{10}),  
$$
where the last two entries are frozen variables 
corresponding to the parameters: 
$$ 
\tx_9=\al, \qquad \tx_{10}=\be ; 
$$
and we must introduce an extended exchange matrix of size $10\times 8$, as follows: 
\begin{equation} \label{extmatD42} 
    \hat{B} = 
 \mqty(0 & 0 & -1 & 0 & 1 & 1 & 0 & 0 \\ 0 & 0 & 1 & 0 & -1 & -1 & 0 & 0 \\ 1 & -1 & 0 & -1 & 0 & 1 & 1 & -1 \\ 0 & 0 & 1 & 0 & -1 & -1 & 0 & 0 \\ -1 & 1 & 0 & 1 & 0 & -1 & -1 & 1 \\-1 & 1 & -1 & 1 & 1 & 0 & -1 & 1 \\ 0 & 0 & -1 & 0 & 1 & 1 & 0 & 0 \\ 0 & 0 & 1 & 0 & -1 & -1 & 0 & 0 \\ -1 & 1 & 0 & 0 & 0 & 0 & 0 & 0  \\ -1 & 0 & 0 & 1 & 0 & 0 & 0 & 0) . 
\end{equation}
The $8\times 8$ submatrix consisting of the first 8 rows of $\hat{B}$ is just the skew-symmetric matrix $\tilde{B}$ determined by the presymplectic form $\tilde\om$, as in \eqref{tom} above, while the last two rows define the arrows that link the frozen nodes (labelled 9 and 10) with the mutable nodes in the associated quiver, as shown in Figure~\ref{quiverD4ii}.

The whole point of this construction is that we can now relate the symplectic map \eqref{D42map} to mutations in the extended cluster algebra 
${\cal A}(\hat{\bx},\hat{B})$. To do this, we consider the following sequence of mutations: 
\begin{equation}\label{D42xeqns}
    \begin{split}
    \tmu_1: \quad&\tx_1'\tx_1 = \tx_5\tx_6\tx_9\tx_{10} + \tx_3, \\ 
  \tmu_3: \quad  &\tx_3'\tx_3 = 
  \tx_1'\tx_2\tx_4\tx_{8}
   + \tx_5(\tx_6)^2\tx_7\tx_9\tx_{10}, \\ 
   \tmu_4:  \quad&\tx_4'\tx_4 = 
  \tx_6\tx_7\tx_{9}
   + \tx_3', \\ 
   \tmu_2: \quad &\tx_2'\tx_2 = 
  \tx_6\tx_7\tx_{10}
   + \tx_3'. 
    \end{split}
\end{equation}
(Note that we denote these mutations by $\tmu_j$, 
in order to distinguish them from mutations 
$\mu_j$ in the original $\rD_4$ cluster algebra.) 
The exchange matrix, and the associated quiver, 
is mutation-periodic with respect to this sequence of mutations, in the sense that their consecutive combination is just equivalent to an overall permutation $\rho$ acting on the nodes, that is 
$$ 
\tmu_2\tmu_4\tmu_3\tmu_1(\hat{B}) = \rho (\hat{B}), \qquad \rho = (24)(18567).
$$
This means that the cluster map 
$$ 
\psi = \rho^{-1}\tmu_2\tmu_4\tmu_3\tmu_1 
$$ 
acts on the cluster $\tilde{\bx}$ as the birational map 
\beq \label{psimap}
\psi: \quad 
(\tx_1,\tx_2,\tx_3,\tx_4,\tx_5,\tx_6,\tx_7,\tx_8) 
\mapsto 
(\tx_8,\tx_4',\tx_3',\tx_2',\tx_6,\tx_7,\tx_1',\tx_5), 
\eeq 
with the primed variables defined by  \eqref{D42xeqns}, 
while it leaves $\hat{B}$ invariant, so $\psi(\hat{B})=\hat{B}$, and hence the presymplectic form \eqref{tom} is preserved, i.e. $\psi^*\,\tilde{\om}=\tilde{\om}$. Then 
it can be verified directly that this cluster map 
is a lift of the symplectic map $\hat{\varphi}$, in the sense that 
$$ 
\hat{\varphi} \cdot \tilde{\pi}=\tilde{\pi} \cdot \psi. 
$$
Furthermore, $\psi$ is a Laurentification of 
$\hat{\varphi}$, in the sense that components of the iterates of $\psi$ are cluster 
variables in the extended cluster algebra, 
and hence belong to the Laurent polynomial ring 
$$
\Z [ \tx_1^{\pm 1}, \tx_2^{\pm 1},\tx_3^{\pm 1},\tx_4^{\pm 1},\tx_5^{\pm 1},\tx_6^{\pm 1},\tx_7^{\pm 1},\tx_8^{\pm 1},\tx_9,\tx_{10} ], 
$$
with the coefficients of the Laurent polynomials 
being polynomials in the frozen variables 
$\tx_9=\al$, $\tx_{10}=\be$. 

\begin{remark}\label{remD4iQ} In 
\cite{hkm24} we also obtained a lift of the type (1) deformed map in Theorem~\ref{defd4} to an extended cluster algebra, whose mutable part 
is given by a seed corresponding to the same 
subquiver with 8 unfrozen nodes (labelled 1-8) as in Figure~\ref{quiverD4ii}. 
However, the two frozen nodes for type (1) are connected to the subquiver differently, and the type (1) symplectic map is the lift of a cluster map built from a sequence of 3 mutations, instead of 4 as in \eqref{D42xeqns}. The mutable subquiver in Figure~\ref{quiverD4ii} 
is also mutation equivalent to the quiver 
found by  Okubo \cite{okubo}, for which a 
certain combination of 
Y-seed mutations 
generates the $q$-Painlev\'{e} VI equation. 
\end{remark} 

\subsection{
The commuting map $\hat\chi$
}

Our starting point for finding a commuting map is the observation that the quiver in Figure~\ref{quiverD4ii}, or equivalently the extended 
exchange matrix $\hat{B}$, 
admits a  symmetry in the form of another sequence of mutations that leaves it invariant up to permutation, namely   
\beq\label{period6}
\tmu_8\tmu_1\tmu_6\tmu_7\tmu_3\tmu_2 (\hat{B}) = \rho^\dagger(\hat{B}), \qquad \mathrm{with}\quad \rho^\dagger 
= (487)(16253). 
\eeq 
Due to this mutation periodicity, 
we have another cluster map, defined by the composition 
$$ 
\chi =(\rho^\dagger)^{-1}\tmu_8\tmu_1\tmu_6\tmu_7\tmu_3\tmu_2. 
$$
The action on the cluster $\tilde{\bx}$ is given in full as  
\beq\label{chimap} 
\chi: 
\quad 
(\tx_1,\tx_2,\tx_3,\tx_4,\tx_5,\tx_6,\tx_7,\tx_8) 
\mapsto 
(\tx_6^\dagger,\tx_5,\tx_1^\dagger,\tx_8^\dagger,\tx_3^\dagger,\tx_2^\dagger,\tx_4,\tx_7^\dagger), 
\eeq 
with the daggered variables defined by  
\begin{equation}\label{D42meqns}
\begin{array}{lrclcl}
 \tmu_2:\,  &  \tx_2^\dagger\,\tx_2 &=&  \tx_5\tx_6\tx_9 &+& \tx_3, \\ 
  \tmu_3:\,  & \tx_3^\dagger\,\tx_3 &=&  \tx_4\tx_5\tx_8\tx_9 &+& 
  \tx_1 \tx_2^\dagger\, \tx_7, \\ 
  \tmu_7:\,  & \tx_7^\dagger\,\tx_7 &= &
  \tx_4\tx_8 \tx_9 &+&\tx_3^\dagger \tx_6, \\
   \tmu_6:\,  & \tx_6^\dagger\,\tx_6 &=& 
   \tx_1\tx_2^\dagger\tx_9 &+&\tx_5\tx_7^\dagger,  \\ 
 \tmu_1:\,   & \tx_1^\dagger\,\tx_1 &=&  
 \tx_3^\dagger \tx_5 \tx_7^\dagger \tx_{10} &+&\tx_4\tx_6^\dagger \tx_8, \\
  \tmu_8:\, &   \tx_8^\dagger\,\tx_8 &=&\tx_2^\dagger \tx_3^\dagger \tx_{10} &+&\tx_{1}^\dagger, 
    \end{array}
\end{equation}
and $\chi(\hat{B})=\hat{B}$ implies that in particular the submatrix $\tilde{B}$ is left invariant, hence $\chi^*\, \tilde{\om}=\tilde{\om}$. 
Observe that the above exchange relations have a natural weighted homogeneity property, if we assign weight 2 to $\tx_3$, weight zero to the frozen variables $\tx_9,\tx_{10}$, and weight 1 to all other $\tx_j$, while $\tx_1^\dagger$ has weight 2 and the other $\tx_j^\dagger$ have weight 1; and  similarly for the exchange relations in \eqref{D42xeqns}.

The main result on the map \eqref{chimap} is the following 
\begin{thm}\label{commuchi}
The cluster map $\chi=(\rho^\dagger)^{-1}\tmu_8\tmu_1\tmu_6\tmu_7\tmu_3\tmu_2$ is the lift of an integrable birational symplectic map 
$\hat\chi$, which acts on the plane with coordinates $(z,w)$ defined by \eqref{pitdef}. It commutes with the 
deformed $\rD_4$ symplectic map $\hat\varphi$ given by \eqref{D42map}, and preserves the same pencil of curves \eqref{Klevel}. 
\end{thm}
\begin{prf}
The proof is by direct calculation. 
First of all, it is necessary to show that  there is a  map 
\beq\label{chihatform}
\hat{\chi}:  
\quad  \mqty(z \\ w) \mapsto 
\mqty(z^\dagger \\ w^\dagger)
\eeq 
such that $\tilde{\pi}\cdot \chi  = 
\hat{\chi} \cdot \tilde{\pi} $, with the quantities
$z^\dagger, w^\dagger$ on the right-hand side being rational functions of the coordinates $z,w$. So from (\ref{tom}) we must find the quantity  
$$\chi^*\left(\frac{\tx_3}{\tx_5\tx_6} \right) =  \frac{\tx_1^\dagger}{\tx_3^\dagger\tx_2^\dagger}, 
$$ 
and then we verify that this is the pullback under 
$\tilde{\pi}$ of the function 
\beq \label{zdag} 
z^\dagger=\frac{\rP(z,w)}{z\,(\al+z) (\al w+\al+z) } 
\eeq 
where 
\begin{align*}
    \rP(z,w) &= \al^{2} \be z w^{2}+\al^{2} \be z w+\al^{2} \be w^{2}+\al^{2} z^{2} w+\al \be z^{2} w+\al z^{3} w+\al z^{2} w^{2} \\
    & +2 \al^{2} \be w+2 \al \be z w+\al z w^{2}+\al^{2} \be+2 \al \be z+\al z w+\be z^{2}+z^{2} w .
\end{align*}
Similarly, for $w^\dagger$ we calculate 
$$ 
\chi^* \left(\frac{\tx_2\tx_4\tx_8}{\tx_1\tx_6\tx_7}\right) =
\frac{\tx_5\tx_8^\dagger\tx_7^\dagger}{\tx_6^\dagger\tx_2^\dagger\tx_4},  
$$
and find that this is the pullback of 
\beq\label{wdag}
w^\dagger =  \frac{(\al w+\al+z)\cdot (\al z w+\al w+\al+z)\cdot (z^\dagger+\be)}{w\, z\,  (\al^{2} z+\al z^{2}+\al z w+\al w+\al+z)} .
\eeq 
Hence we have a well defined  map of the plane, 
and the fact  that $\hat\chi$ is birational essentially follows from the birationality of 
the cluster map $\chi$, which is composed of mutations (which are involutions) together with a permutation.  Now we can also check 
directly that 
$\hat{\varphi}\cdot\hat{\chi} = 
\hat{\chi}\cdot\hat{\varphi}$, 
so the maps commute, and by lifting 
we find 
that the cluster maps $\psi$ and $\chi$ also commute with one another, i.e. 
\beq\label{cmaps}{\psi}\cdot{\chi} = 
{\chi}\cdot{\psi} . 
\eeq  
Furthermore, 
the fact that $\chi$ preserves the 2-form \eqref{tom} implies that $\hat{\chi}$ preserves 
$\hat\om$, so $\hat{\chi}$ is symplectic, and 
for the function \eqref{D42firstint} it can further be verified that
$\hat{\chi}^* ( K ) =K$. Hence $\hat{\chi}$
is an integrable map, and it preserves the same 
pencil \eqref{Klevel} as $\hat{\varphi}$ 
does. 
\end{prf}

\begin{remark}\label{undefchi}
In the undeformed case when $\al=1$, $\be=1$, 
the above expressions 
\eqref{zdag} and \eqref{wdag} 
for $z^\dagger,w^\dagger$ in \eqref{chihatform} simplify considerably, and the new map $\hat\chi$ reduces to the old one, so $\hat{\chi}=\hat{\varphi}$.  
\end{remark}

\subsection{Tau functions on the $\Z^2$ lattice}

The process of Laurentification, lifting a birational map to a cluster map for which the Laurent property holds, is analogous to the process of introducing tau functions in  integrable systems, and in some cases (namely, for discrete Hirota equations), the two things coincide. Hence, in the case at hand, we will refer to the sequences of cluster variables obtained under the actions of the cluster maps $\psi$ and $\chi$ as tau functions.

\begin{figure}[h]
\centering
\epsfig{file={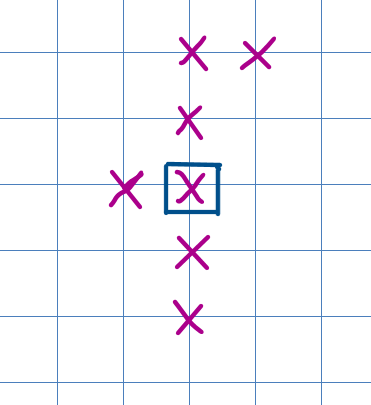}, height=1.8in, width=1.8in}
\caption{Initial stencil of  tau functions on $\Z^2$. }
\label{taustencil}
\end{figure}

Due to the commutativity of the two cluster maps, as in 
(\ref{cmaps}), we can combine their action to get a set of clusters defined by tau functions indexed by $(m,n)\in\Z^2$, by taking the initial cluster to be 
\beq\label{TXseed}
\tilde{\bx} = (T_{0,-2},T_{-1,0},X_{0,0},T_{1,2},T_{0,0},T_{0,1},T_{0,2},T_{0,-1})= 
(\tx_1, \tx_2,\tx_3,\tx_4,\tx_5, \tx_6,\tx_7,\tx_8) ,
\eeq 
and then defining 
\beq 
\label{TXmn}
\tilde{\bx}_{m,n} = 
\chi^m\psi^n (\tilde{\bx} )
= 
(T_{m,n-2},T_{m-1,n},X_{m,n},T_{m+1,n+2},T_{m,n},T_{m,n+1},T_{m,n+2},T_{m,n-1}).
\eeq 
The initial cluster \eqref{TXseed} corresponds to the stencil of lattice points shown in Figure~\ref{taustencil}, 
where the 7 crosses represent the tau functions $T_{ij}$, and the box represents $X_{0,0}$, placed at the origin in $\Z^2$; and 
each cluster \eqref{TXmn} corresponds to the same stencil but shifted $m$ steps to the right/left and $n$ steps up/down, according to whether $m,n$ are positive/negative, respectively. 
Note that, due to the different weights of the equations 
\eqref{D42meqns}, we are required to use two different tau functions on the lattice: $T_{m,n}$, which has weight 1, and $X_{m,n}$, which has weight 2. By considering the action of 
 the symplectic maps $\hat{\chi}$, $\hat{\varphi}$ on the plane with coordinates 
 $\bu =(z,w)$,
we can also define a corresponding sequence of 
combined iterates of these two maps, 
denoted by 
\beq\label{yseqmn}
\hat{\chi}^m\hat{\varphi}^n (\bu ) = (z_{m,n},w_{m,n}), \qquad 
\mathrm{with}\quad
z_{m,n}=\frac{X_{m,n}}{T_{m,n}T_{m,n+1}}, \quad 
w_{m,n}=\frac{T_{m-1,n}T_{m+1,n+2}T_{m,n-1}}{T_{m,n-2}T_{m,n+1}T_{m,n+2}},
\eeq 
where $\bu\in\C^2$ is an arbitrary initial point in the affine plane, 
corresponding to the seed (\ref{TXseed}). 
This is in accordance with the map \eqref{pitdef}, sending 
$$ 
\tilde{\pi}: \qquad \tilde{\bx}_{m,n} \mapsto 
\hat{\chi}^m\hat{\varphi}^n (\bu ) . 
$$

\begin{thm}\label{taufnTX}
The tau functions on $\Z^2$, associated via (\ref{yseqmn}) with combined iteration of the commuting 
integrable maps $\hat\chi$ and $\hat\varphi$,
satisfy the following system of weighted homogeneous equations:
\begin{align}
T_{m,n+3}T_{m,n-2}
& =  \al\be\, T_{m,n+1}T_{m,n} & +\quad&  X_{m,n}, \refstepcounter{equation}\subeqn\label{hirotaeqs1}\\ 
X_{m,n+1}X_{m,n} & =  T_{m+1,n+2} T_{m,n+3}T_{m,n-1}T_{m-1,n}
& +\quad&  \al\be\, T_{m,n+2}T_{m,n+1}^2T_{m,n},  \subeqn\label{hirotaeqs2}\\ 
T_{m+1,n+2}T_{m-1,n+1}  & =  \al\, T_{m,n+2}T_{m,n+1} &+\quad  &X_{m,n+1}, \subeqn\label{hirotaeqs3}\\ 
T_{m+1,n+3}T_{m-1,n}  & =  \be\, T_{m,n+2}T_{m,n+1} &+\quad  &X_{m,n+1}, \subeqn\label{hirotaeqs4}\\ 
T_{m+1,n}X_{m,n} & =  \al\,T_{m+1,n+2}  T_{m,n-1}T_{m,n}& +\quad & T_{m+1,n+1}T_{m,n+2}T_{m,n-2},\subeqn\label{hirotaeqs5}\\
T_{m+1,n-1}T_{m,n+2} & =  \al\, T_{m,n-1}T_{m+1,n+2} &+\quad &T_{m,n+1}T_{m+1,n}, \subeqn\label{hirotaeqs6}\\
X_{m+1,n}T_{m,n-2} & =  \be\,T_{m+1,n} T_{m+1,n-1} T_{m,n} & +\quad & T_{m+1,n+2}T_{m+1,n-2}T_{m,n-1}
.\subeqn\label{hirotaeqs7}
\end{align}
Conversely, any solution of this system of bilinear lattice equations produces a simultaneous solution $(z_{m,n}, w_{m,n})$ 
of the pair of iterated maps  $\hat\chi$ and $\hat\varphi$. 
\end{thm}
\begin{prf} At any fixed lattice point $(m,n)\in\Z^2$, by starting from a seed cluster $\tilde{\bx}_{m,n}$ given by \eqref{TXmn}, corresponding to the points in the stencil shown in Figure~\ref{taustencil}, we can identify the cluster variables via   
$$
\tilde{\bx}_{m,n}=(\tx_1, \tx_2,\tx_3,\tx_4,\tx_5, \tx_6,\tx_7,\tx_8),
$$ 
just as in \eqref{TXseed} for the case $m=0=n$, and further extend this cluster by adjoining the frozen variables 
$\tx_9=\al, \tx_{10}=\be$. To compute the action of 
$\chi$ on this cluster, we begin by applying the mutation 
$\tmu_2$, which corresponds to the equation \eqref{hirotaeqs3} after shifting down the index $n\to n-1$, and produces the new cluster variable 
$\tx_2^\dagger=T_{m+1,n+1}$. Subsequently, we apply (in order) the mutations $\tmu_3,\tmu_7,\tmu_6,\tmu_1,\tmu_8$, corresponding to the equations  \eqref{hirotaeqs5}, 
\eqref{hirotaeqs6}, \eqref{hirotaeqs6} again but with  $n\to n-1$, \eqref{hirotaeqs7}, \eqref{hirotaeqs4} 
 but with  $m\to m+1$ and $n\to n-1$, which produce the new cluster variables 
$\tx_3^\dagger=T_{m+1,n}$, $\tx_7^\dagger=T_{m+1,n-1}$, 
$\tx_6^\dagger=T_{m+1,n-2}$, $\tx_1^\dagger=X_{m+1,n}$, $\tx_8^\dagger=T_{m+2,n+2}$, respectively. Finally, we apply 
the permutation $(\rho^\dagger)^{-1}$, which from 
\eqref{chimap} yields 
$$ 
\chi (\tilde{\bx}_{m,n} )
= 
(T_{m+1,n-2},T_{m,n},X_{m+1,n},T_{m+2,n+2},T_{m+1,n},T_{m+1,n+1},T_{m+1,n+2},T_{m+1,n-1}) = \tilde{\bx}_{m+1,n}, 
$$ 
that is, the new seed with $m\to m+1$, which corresponds to shifting all the crosses and the box that make up the stencil in Figure~\ref{taustencil} one place to the right. 
Moreover, by the argument used to prove Theorem \ref{commuchi}, the pair of quantities $(z_{m+1,n},w_{m+1,n})$ given by the appropriate ratios of these cluster variables, as defined by the expressions for $z_{m,n}$ and $w_{m,n}$ in \eqref{yseqmn} with 
$m\to m+1$, are precisely the image of 
$(z_{m,n},w_{m,n})$ under the map $\hat\chi$, that is 
$$ 
\hat\chi \Big( (z_{m,n},w_{m,n}) \Big) = (z_{m+1,n},w_{m+1,n}). 
$$

Similarly, starting from the same seed cluster  
$\tilde{\bx}_{m,n}$ with the two frozen variables $\al,\be$, we can compute the action of the cluster map $\psi$, as in \eqref{psimap}. Then we see that taking the mutations 
$\tmu_1,\tmu_3,\tmu_4,\tmu_2$, in that order, corresponds to the equations  \eqref{hirotaeqs1}, 
\eqref{hirotaeqs2}, \eqref{hirotaeqs3},  \eqref{hirotaeqs4}, 
which produce the new cluster variables 
$\tx_1'=T_{m,n+3}$, $\tx_3'=X_{m,n+1}$, $\tx_4'=T_{m-1,n+1}$, $\tx_2'=T_{m+1,n+3}$, respectively. Thus, after applying the permutation $\rho^{-1}$, we obtain the new cluster, namely 
$$ 
\psi (\tilde{\bx}_{m,n} )
= 
(T_{m,n-1},T_{m-1,n+1},X_{m,n+1},T_{m+1,n+3},T_{m,n+1},T_{m,n+2},T_{m,n+3},T_{m,n}) = \tilde{\bx}_{m,n+1}, 
$$ 
which is the new seed with $n\to n+1$, and this corresponds to moving all the crosses and the box in Figure~\ref{taustencil} one place up. The result of Theorem 5.6 in \cite{hkm24} also shows that suitable  ratios of the tau functions from the above cluster produce the pair  $(z_{m,n+1},w_{m,n+1})$, given by applying the shift $n\to n+1$ to the formulae for $z_{m,n}$ and $w_{m,n}$ in \eqref{yseqmn}, and this coincides with the 
image of 
$(z_{m,n},w_{m,n})$ under the symplectic map $\hat\varphi$, that is
$$ 
\hat\varphi \Big( (z_{m,n},w_{m,n}) \Big) = (z_{m,n+1},w_{m,n+1}). 
$$

To complete the proof, it is enough to note the fact that, from the commutativity of the two cluster maps, as in \eqref{cmaps}, it is consistent to extend the system 
\eqref{hirotaeqs1}-\eqref{hirotaeqs7} to all $(m,n)\in\Z^2$, and then from \eqref{yseqmn} the solution of this lattice system provides a simultaneous solution for the pair of iterated commuting maps $\hat\chi$ and $\hat\varphi$.
\end{prf}

\subsection{Degree growth of cluster maps}

The Laurent property means that each of the tau functions $T_{m,n},X_{m,n}$ satisfying the lattice system \eqref{hirotaeqs1}-\eqref{hirotaeqs7} is a Laurent polynomial in the initial seed variables \eqref{TXseed} with coefficients in $\Z[\al,\be]$. Hence we can write each of these cluster variables in the form 
\beq\label{lpolys}
T_{m,n} = \frac{\rN_{m,n}(\hat{\bx})}{\tilde{\bx}^{\bd_{m,n}}}, \qquad 
X_{m,n} = \frac{\rN^*_{m,n}(\hat{\bx})}{\tilde{\bx}^{\boe_{m,n}}}, 
\eeq 
where $\rN_{m,n},\rN^*_{m,n}\in \Z[\hat{\bx}]$ are polynomials in the extended cluster variables $\hat{\bx}$ that are not divisible by any $\tx_j$, $1,\leq j\leq 8$, and the denominators are monomials in the seed variables $\tilde{\bx}$, with the exponent of each $\tx_j$ being specified by the $j$th component of the d-vectors $\bd_{m,n}, \boe_{m,n}\in \Z^8$, respectively. Due to the homogeneity properties of these tau functions, as observed above, the degrees of each numerator and denominator are completely determined by the d-vectors.  Then, in order to calculate the growth of degrees, we can use the basic fact, pointed out in \cite{fz4}, that the d-vectors for cluster variables satisfy the tropical (max-plus) version of the exchange relations, in which the frozen variables do not make any contribution.  This immediately yields the following result, by taking the appropriate max-plus analogues  of \eqref{hirotaeqs1}-\eqref{hirotaeqs7}.  

\begin{propn}\label{trop}
The d-vectors for the tau functions \eqref{lpolys} on the $\Z^2$ lattice satisfy the max-plus equations 
\beq\label{trophirotaeqs}
\begin{array}{rcl}
\bd_{m,n+3}+\bd_{m,n-2}
& = & \max( \bd_{m,n+1}+\bd_{m,n} ,  \boe_{m,n}), \\ 
\boe_{m,n+1}+\boe_{m,n} & = & \max( \,\bd_{m+1,n+2}+ \bd_{m,n+3}+\bd_{m,n-1}+\bd_{m-1,n}
, \, \bd_{m,n+2}+2\bd_{m,n+1}+\bd_{m,n}),  \\ 
\bd_{m+1,n+2}+\bd_{m-1,n+1}  & = & \max ( \,\bd_{m,n+2}+\bd_{m,n+1} ,\, \boe_{m,n+1}), \\ 
\bd_{m+1,n+3}+\bd_{m-1,n}  & = & \max( \,\bd_{m,n+2}+\bd_{m,n+1} , \,\boe_{m,n+1}), \\ 
\bd_{m+1,n}+\boe_{m,n} & = & \max(\,\bd_{m+1,n+2}  +\bd_{m,n-1}+\bd_{m,n}, \,\bd_{m+1,n+1}+\bd_{m,n+2}+\bd_{m,n-2}),\\
\bd_{m+1,n-1}+\bd_{m,n+2} & = & \max(\, \bd_{m,n-1}+\bd_{m+1,n+2} ,\,\bd_{m,n+1}+\bd_{m+1,n}), \\
\boe_{m+1,n}+\bd_{m,n-2} & = & \max(\,\bd_{m+1,n} +\bd_{m+1,n-1}+ \bd_{m,n} , \,\bd_{m+1,n+2}+\bd_{m+1,n-2}+\bd_{m,n-1})
.
\end{array}
\eeq 
\end{propn} 

Although the system of max-plus equations \eqref{trophirotaeqs} appears rather complicated, the presence of Zamolodchikov periodicity in the original $\rD_4$ cluster map entails that there are special combinations of d-vectors that are periodic in both $m$ and $n$, and this means the system can be linearized, and thereby solved exactly. For our purposes, we are only interested in a specific set of initial conditions, corresponding to the seed \eqref{TXseed}. Specifically, this means we have a tropical seed for the max-plus relations, given by the matrix 
\beq\label{tropseed}
\tilde{\bx} = (\bd_{0,-2},\bd_{-1,0},\boe_{0,0},\bd_{1,2},\bd_{0,0},\bd_{0,1},\bd_{0,2},\bd_{0,-1})= 
-I_8 ,
\eeq 
where $I_8$ denotes the $8\times 8$ identity matrix. 

The particular combinations of d-vectors that we need to consider are those given by the tropical 
versions of the formulae in \eqref{yseqmn}, which express the coordinate pair $(z_{m,n},w_{m,n})$ in terms of the tau functions, namely 
\beq\label{troppi}
\bZ_{m,n} = \boe_{m,n} -\bd_{m,n}-\bd_{m,n+1}, 
\qquad \bW_{m,n} = \bd_{m-1,n}+\bd_{m+1,n+2} 
+\bd_{m,n-1} - \bd_{m,n-2}- \bd_{m,n+1}- \bd_{m,n+2}. 
\eeq 
The main point is that 
the pair $(\bZ_{m,n},\bW_{m,n})$ satisfies 
max-plus versions of the maps $\hat\varphi$ and $\hat\chi$. 

\begin{lem}\label{tropmaps}
Let $\bV_{m,n}=[\bZ_{m,n}]_+$. Then under the action of the shift $n\mapsto n+1$,  the pair $(\bZ_{m,n},\bW_{m,n})$ is transformed by the max-plus version of the map \eqref{newphihatmapD4} defined by 
$$ 
\hat{\varphi}_{\mathrm{trop}}: \quad 
(\bZ_{m,n},\bW_{m,n}) \mapsto(\bZ_{m,n+1},\bW_{m,n+1}), 
$$
where 
\beq\label{phitrop}
\begin{array}{rcl} 
\bZ_{m,n+1} + \bZ_{m,n} 
& = &  [\bV_{m,n}+\bW_{m,n}]_+, \\ 
\bW_{m,n+1} + \bW_{m,n} 
& = &  2[\bW_{m,n}]_+ + \bV_{m,n}-2 \bZ_{m,n} .
\end{array}
\eeq 
The action of the shift $m\mapsto m+1$ is  
the max-plus version of \eqref{chihatform} given by 
$$ 
\hat{\chi}_{\mathrm{trop}}: \quad 
(\bZ_{m,n},\bW_{m,n}) \mapsto(\bZ_{m+1,n},\bW_{m+1,n}), 
$$
where 
\beq\label{chitrop}
\begin{array}{rcl} 
\bZ_{m+1,n} + \bZ_{m,n} 
& = &  [\bV_{m,n}+\bW_{m,n}]_+, \\ 
\bW_{m+1,n} + \bW_{m,n} 
& = &  2[\bW_{m,n}]_+ + \bV_{m,n}-2 \bZ_{m,n} ,
\end{array}
\eeq 
which, considered as a map on pairs of vectors, is isomorphic to 
$\hat{\varphi}_{\mathrm{trop}}$. 
Given arbitrary initial values $ (Z_{0,0}, W_{0,0})\in \R^2$, every component of the iterates of either map is periodic with
period 4.
\end{lem}
\begin{prf} Observe that, by subtracting 
$\bd_{m,n}+\bd_{m,n+1}$ 
from both sides of the first equation in the system \eqref{trophirotaeqs}, and using the definition of $\bZ_{m,n}$ in \eqref{troppi},  the quantity $\bV_{m,n}=[\bZ_{m,n}]_+$ is given in terms of d-vectors by 
\beq\label{Vdef}
\bV_{m,n} = \bd_{m,n+3} - \bd_{m,n+1}-\bd_{m,n} +\bd_{m,n-2}. 
\eeq 
By rearranging this and the other equations in the max-plus system that correspond to the shift $n\mapsto n+1$, one finds that the pair 
$(\bZ_{m,n},\bW_{m,n})$ is transformed according to the max-plus version of $\hat\varphi$ given in 
\eqref{phitrop}: this is just the content of Lemma 5.9 in \cite{hkm24}, where it was also 
noted that every orbit of this map is periodic with period 4, corresponding to the Zamolodchikov period of the original $\rD_4$ map, so that $(\bZ_{m,n+4},\bW_{m,n+4})= (\bZ_{m,n},\bW_{m,n}) $. Now, because 
the tropicalization resulting in the d-vector equations is blind to the coefficients $\al,\be$, it follows that the relevant max-plus equations for the action of the shift $m\mapsto m+1$ on $(\bZ_{m,n},\bW_{m,n})$ are, in particular, given by the tropical version of $\hat\chi$ when 
$\al=\be=1$. However, as pointed out in Remark~\ref{undefchi}, in this case the two symplectic maps coincide, hence their tropical versions must be the same, so that $\hat{\chi}_{\mathrm{trop}}$ is given by \eqref{chitrop}, 
and we have the same period 4 behaviour in $m$: 
$(\bZ_{m+4,n},\bW_{m+4,n})=(\bZ_{m,n},\bW_{m,n})$. 
\end{prf}

In order to analyse the growth of degrees it is convenient to introduce the shift operators $\cal S$, $\cal T$ for the shifts $m\mapsto m+1$ and 
$n\mapsto n+1$, respectively, which act on any function of $m$ and $n$ according to  ${\cal S} f_{m,n}=f_{m+1,n}$, ${\cal T} f_{m,n}=f_{m,n+1}$. 

\begin{thm}\label{degrowth} 
The d-vectors $\bd_{m,n}$, $\boe_{m,n}$ lie in the kernel of the linear shift operators 
$$ {\cal L}^\dagger =  
({\cal S}^4-1)({\cal S}^3-1)({\cal S}-1), 
\qquad \tilde{\cal L} 
=({\cal T}^4-1)({\cal T}^3-1)({\cal T}^2-1),
$$
as well as satisfying mixed linear relations,  such as 
\beq\label{mixed}
({\cal S}^4-1)({\cal T}^3-1)({\cal T}^2-1)\, \bd_{m,n}=0. 
\eeq 
Subject to the initial conditions 
specified by the seed \eqref{tropseed},  
the leading order growth of the d-vectors 
as $m,n\to\infty$ is given by 
\beq\label{lorder} 
\bd_{m,n}= \boa\, n^2 +\bob \,nm +\boc\, m^2 + O(n) + O(m), \qquad 
\boe_{m,n}\sim 2 \bd_{m,n}, 
\eeq 
where 
\begin{align*}
    \boa = \frac{1}{24} (1,1,2,1,1,1,1,1)^{T} , \qquad \bob=-2\boa, \qquad \boc = 7\boa . 
\end{align*}
\end{thm}
\begin{prf}
The $n$ part of the statement, namely the fact that the d-vectors lie in the kernel of 
$\tilde{\cal L}$,  was already given in Theorem 5.11 of \cite{hkm24}, 
together with the  coefficient vector $\boa$ multiplying the leading $n^2$ growth in this lattice direction. The proof of the rest of the statement follows a  very similar pattern, and also closely resembles the proofs of analogous results on the solution of a tropical system associated with a deformed $\rA_3$ map 
(cf.\ Lemma 2.10 and Theorem 2.11 in \cite{hkm24}). The main point is that the solution of \eqref{trophirotaeqs} on $\Z^2$ is completely determined by fixing a tropical seed, with the specific choice \eqref{tropseed} being the one that corresponds to \eqref{TXseed}.  Then we see that, 
on the one hand,  the expression \eqref{Vdef} can be rewritten as 
\beq\label{Vd}
\bV_{m,n} = ({\cal T}^3-1)({\cal T}^2-1)\, \bd_{m,n-2}, 
\eeq 
while on the other hand, by the 4-periodicity in Lemma~\ref{tropmaps}, we have 
$$ 
({\cal T}^4-1)\, \bZ_{m,n}=0 = 
({\cal S}^4-1)\, \bZ_{m,n}
\implies 
({\cal T}^4-1)\, \bV_{m,n}=0 = 
({\cal S}^4-1)\, \bV_{m,n}, 
$$
which together with \eqref{Vd} imply that the linear relations 
$\tilde{\cal L}\, \bd_{m,n}=0$ and  \eqref{mixed} 
both hold. To see that the other d-vector satisfies these same linear relations, it is sufficient to observe from \eqref{troppi} that 
\beq\label{We}
\boe_{m,n} = \bZ_{m,n} + \bd_{m,n} + \bd_{m,n+1}, 
\eeq 
and then apply in turn the operators $\tilde{\cal L}$ and $({\cal S}^4-1)({\cal T}^3-1)({\cal T}^2-1)$ to both sides of the above equation, using the fact that 
$({\cal T}^4-1)\, \bZ_{m,n}=0 = 
({\cal S}^4-1)\, \bZ_{m,n}$. Now the linear relation for ${\cal L}^\dagger$ is obtained by considering the equations \eqref{trophirotaeqs} in turn, and first eliminating $\boe_{m,n}$ and shifts in $n$, in order to obtain 
${\cal L}^\dagger\, \bd_{m,n}=0$ (for arbitrary fixed $n$), and the preceding argument with \eqref{We} can be repeated to show that $\boe_{m,n}$ lies in the kernel of the same operator. Given the operators ${\cal L}^\dagger$ and $\tilde{\cal L}$, 
one can then use the initial conditions \eqref{tropseed} to find the exact solution of the corresponding two homogeneous linear equations, in $m$ and $n$ respectively, which is compatible with the overall evolution of the equations \eqref{trophirotaeqs} on $\Z^2$, and this yields not only the quadratic leading order terms in \eqref{lorder}, but also explicit  corrections of order $m$ and $n$, as well as periodic terms, if so desired. It can then be noted from \eqref{We} that, once the growth of $\bd_{m,n}$ has been determined, the 4-periodicity means that $\bZ_{m,n}$ is $O(1)$, hence 
$$ 
\boe_{m,n} \sim \bd_{m,n} + \bd_{m,n+1} \sim 
2 \bd_{m,n}, 
$$
as required. 
\end{prf}

\section{Elliptic surface and degree growth for deformed D$_4$ map}
As noted above, a generic point in 
 $\mathbb{P}^1 \times \mathbb{P}^1$ uniquely determines a value of the first integral
\eqref{D42firstint}, hence specifying a particular curve in  the pencil \eqref{Klevel} which passes through that point. However, this is not the case for the base points: these are the points through which every curve in the pencil passes, and they correspond to singularities of the maps $\hat\varphi$ and $\hat\chi$. In this section we describe how to resolve these singularities and obtain a smooth surface, denoted ${\cal X}_0$. We then use this to calculate the degree growth of $\hat\varphi$ and $\hat\chi$, via their induced action on the Picard lattice. 

\subsection{Space of initial conditions}

Let $(z, w)$ be affine 
coordinates for $\mathbb{P}^1 \times \mathbb{P}^1$ and let $Z = \frac{1}{z}$ and $W = \frac{1}{w}$.
Then, $\mathbb{P}^1 \times \mathbb{P}^1$ is covered by $4$ charts: $(z, w)$, $(Z, w)$, $(z, W)$ and $(Z, W)$.

\begin{figure}
	\begin{picture}(250, 360)
		{\thicklines
		\put(0, 30){\line(1, 0){100}}
		\put(0, 90){\line(1, 0){100}}
		\put(20, 10){\line(0, 1){100}}
		\put(80, 10){\line(0, 1){100}}
		}
		
		\put(20, 60){\circle*{5}}
		\put(0, 57){(1)}
		
		\put(30, 30){\circle*{5}}
		\put(22, 16){(2)}
		
		\put(45, 30){\circle*{5}}
		\put(37, 16){(3)}
		
		\put(60, 30){\circle*{5}}
		\put(52, 16){(4)}
		
		\put(30, 90){\circle*{5}}
		\put(22, 98){(5)}

		\put(50, 90){\circle*{5}}
		\put(42, 98){(6)}

		\put(80, 60){\circle*{5}}
		\put(85, 57){(7)}
		
		\put(80, 90){\circle*{5}}
		\put(85, 98){(8)}

		\put(50, 120){$\downarrow$}

		{\thicklines
		\put(0, 160){\line(1, 0){100}}
		\put(80, 140){\line(0, 1){50}}
		
		\put(20, 140){\line(0, 1){100}}
		\put(0, 220){\line(1, 0){50}}
		\qbezier(80, 190)(80, 205)(100, 220)
		\qbezier(50, 220)(65, 220)(80, 240)
		}
		
		\put(10, 190){\line(1, 0){20}}

		\put(30, 150){\line(0, 1){20}}

		\put(45, 150){\line(0, 1){20}}

		\put(60, 150){\line(0, 1){20}}

		\put(30, 210){\line(0, 1){20}}

		\put(50, 210){\line(0, 1){20}}

		\put(70, 190){\line(1, 0){20}}

		\put(100, 200){\line(-1, 1){40}}

		\put(79, 221){\circle*{5}}
		\put(67, 205){(9)}

		\put(50, 250){$\downarrow$}

		{\thicklines
		\put(0, 290){\line(1, 0){100}}
		\put(80, 270){\line(0, 1){50}}
		
		\put(20, 270){\line(0, 1){100}}
		\put(0, 350){\line(1, 0){50}}
		\qbezier(80, 320)(80, 335)(100, 350)
		\qbezier(50, 350)(65, 350)(80, 370)
		}
		
		\put(10, 320){\line(1, 0){20}}

		\put(30, 280){\line(0, 1){20}}

		\put(45, 280){\line(0, 1){20}}

		\put(60, 280){\line(0, 1){20}}

		\put(30, 340){\line(0, 1){20}}

		\put(50, 340){\line(0, 1){20}}

		\put(70, 320){\line(1, 0){20}}

		\put(100, 330){\line(-1, 1){40}}

		\put(70, 340){\line(1, 1){20}}
		
		\put(100, 300){${\cal X}_1$}

		{\thicklines
		\put(190, 160){\line(1, 0){50}}
		\put(220, 140){\line(0, 1){50}}
		
		\qbezier(220, 190)(220, 205)(240, 220)
		\qbezier(190, 220)(205, 220)(220, 240)

		\qbezier(190, 160)(160, 160)(150, 200)
		\qbezier(190, 220)(160, 220)(150, 180)
		}
		
		\put(143, 190){\line(1, 0){20}}
		\put(153, 190){\circle*{5}}

		\put(170, 155){\line(0, 1){20}}

		\put(185, 150){\line(0, 1){20}}

		\put(200, 150){\line(0, 1){20}}

		\put(170, 205){\line(0, 1){20}}

		\put(190, 210){\line(0, 1){20}}

		\put(210, 190){\line(1, 0){20}}

		\put(240, 200){\line(-1, 1){40}}

		\put(210, 210){\line(1, 1){20}}
		
		\put(120, 250){$\searrow$}

		\put(185, 140){${\cal X}_0$}

	\end{picture}
	\caption{
		The surface ${\cal X}_1$ obtained by blowing up the indeterminacies of the pencil defined by $K$ and the minimal space of initial conditions ${\cal X}_0$.
	}
	\label{figure:blowup}
\end{figure}
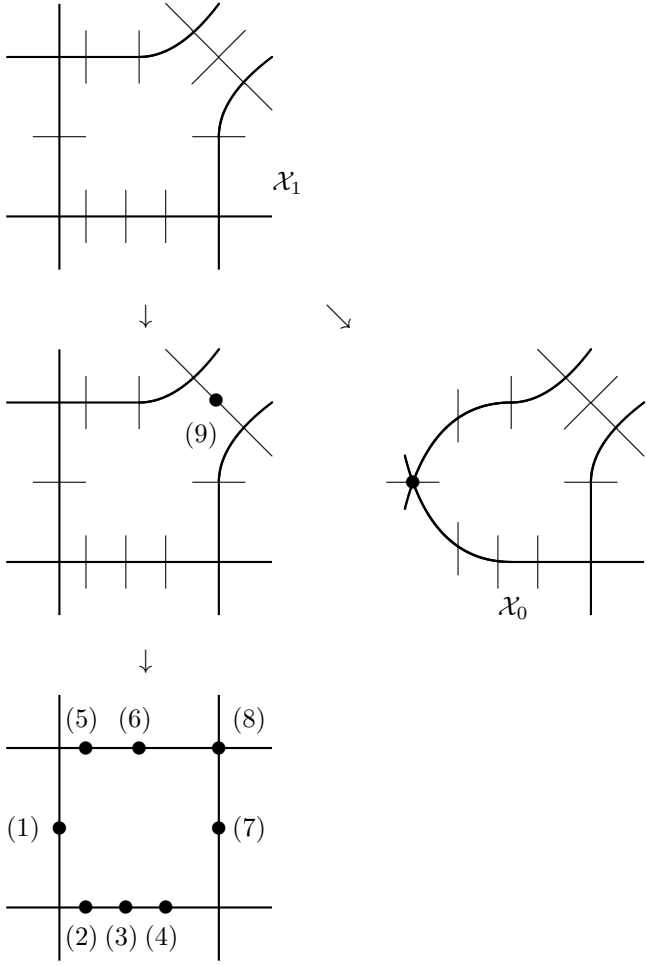

Let us blow up the base points of the pencil determined by the invariant $K$ in  \eqref{D42firstint}, 
which on a fixed level $K=\ka$ is given by 
\eqref{Klevel}. 
The pencil has $9$ base points on $\mathbb{P}^1 \times \mathbb{P}^1$:
\begin{itemize}
	\item[(1)]
	$(z, w) = \left( 0, - 1 \right)$,

	\item[(2)]
	$(z, w) = (- 1, 0)$,

	\item[(3)]
	$(z, w) = (- \alpha, 0)$,

	\item[(4)]
	$(z, w) = (- \beta, 0)$,

	\item[(5)]
	$(z, W) = (- 1, 0)$,

	\item[(6)]
	$(z, W) = (- \alpha \beta, 0)$,

	\item[(7)]
	$(Z, w) = (0, - 1)$,

	\item[(8)]
	$(Z, W) = (0, 0)$,

	\item[(9)]
	$\left( Z, \frac{W}{Z} \right) = (0, - 1)$,

\end{itemize}
where point (9) is infinitely near (8).
Let ${\cal X}_1$ be the surface obtained by blowing up $\mathbb{P}^1 \times \mathbb{P}^1$ at these $9$ points (see Figure~\ref{figure:blowup}).
Then, the pencil has no indeterminacy in ${\cal X}_1$.

However, since the Picard number of ${\cal X}_1$ is $11$ ($> 10$), ${\cal X}_1$ is not minimal as a rational elliptic surface and, in fact, the commuting maps have an indeterminacy on ${\cal X}_1$.
Since it is an exceptional curve of the first kind and contained in a fibre, the strict transform of the curve $\{ z = 0 \}$ can be contracted.
Let ${\cal X}_0$ be the surface obtained after blowing down this curve (see Figure~\ref{figure:blowup}).
Then, ${\cal X}_0$ is the minimal space of initial conditions for the commuting maps.
That is, the maps $\hat{\varphi}$ and $\hat{\chi}$ act on ${\cal X}_0$ as automorphisms.

\begin{figure}
	\begin{picture}(200, 500)
		{\thicklines
		\put(0, 60){\line(1, 0){200}}
		\put(160, 20){\line(0, 1){100}}
		
		\put(40, 20){\line(0, 1){200}}
		\put(0, 180){\line(1, 0){100}}
		\qbezier(160, 120)(160, 150)(200, 180)
		\qbezier(100, 180)(130, 180)(160, 220)
		}
		
		\put(20, 120){\line(1, 0){40}}

		\put(60, 40){\line(0, 1){40}}

		\put(90, 40){\line(0, 1){40}}

		\put(120, 40){\line(0, 1){40}}

		\put(60, 160){\line(0, 1){40}}

		\put(100, 160){\line(0, 1){40}}

		\put(140, 120){\line(1, 0){40}}

		\put(200, 140){\line(-1, 1){80}}

		\put(140, 160){\line(1, 1){40}}
		
		\put(141, 199){\circle*{5}}
		\put(117, 196){(10)}

		\put(179, 161){\circle*{5}}
		\put(185, 158){(11)}

		\put(160, 60){\circle*{5}}
		\put(163, 49){(12)}

		{\thicklines
		\put(0, 320){\line(1, 0){100}}
		
		\put(40, 280){\line(0, 1){200}}
		\put(0, 440){\line(1, 0){100}}
		\qbezier(160, 380)(160, 350)(200, 320)
		\qbezier(100, 320)(130, 320)(160, 280)

		\qbezier(160, 380)(160, 410)(200, 410)
		\qbezier(100, 440)(130, 440)(130, 480)
		}
		
		\put(0, 445){$D_1$}
		\put(145, 485){$D_2$}
		\put(180, 315){$D_3$}
		\put(0, 325){$D_4$}

		\put(25, 280){$E_1$}
		\put(105, 470){$E_2$}
		\put(185, 445){$E_3$}
		\put(190, 365){$E_4$}

		\put(20, 380){\line(1, 0){40}}
		\put(65, 375){$C_1$}

		\put(60, 300){\line(0, 1){40}}
		\put(55, 345){$C_8$}

		\put(90, 300){\line(0, 1){40}}
		\put(85, 345){$C_7$}

		\put(120, 300){\line(0, 1){40}}
		\put(115, 345){$C_6$}

		\put(60, 420){\line(0, 1){40}}
		\put(55, 410){$C_2$}

		\put(100, 420){\line(0, 1){40}}
		\put(95, 410){$C_3$}

		\put(140, 380){\line(1, 0){40}}
		\put(125, 375){$C_5$}

		\qbezier(200, 430)(150, 430)(150, 480)

		\put(140, 420){\line(1, 1){40}}
		\put(130, 410){$C_4$}
		
		\put(120, 280){\line(1, 1){80}}

		\put(190, 400){\line(0, 1){40}}

		\put(120, 470){\line(1, 0){40}}

		\put(100, 250){$\downarrow$}

	\end{picture}
	\caption{
		Blow-ups needed to obtain the surface ${\cal X}$ (upper) from ${\cal X}_1$ (lower).
	}
	\label{figure:surface}
\end{figure}
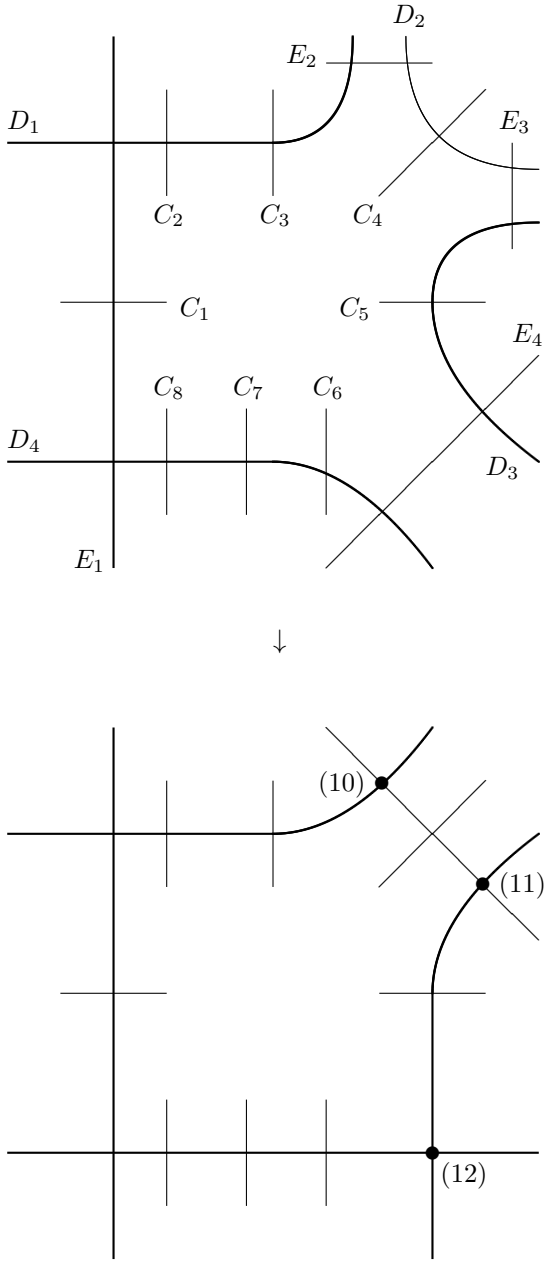

However, the minimization process contracts a curve in the original $\mathbb{C}^2$.
To calculate the degree growth in the original chart, the minimal surface ${\cal X}_0$ is not very convenient.
Therefore, we blow up the surface ${\cal X}_1$ at the locus of the curve $\{ z = 0 \}$ instead of blowing it down.
That is, we blow up ${\cal X}_1$ at the following $3$ points:
\begin{itemize}
	\item[(10)]
	$\left( Z, \frac{W}{Z} \right) = (0, 0)$,

	\item[(11)]
	$\left( \frac{Z}{W}, W \right) = (0, 0)$,

	\item[(12)]
	$(Z, w) = (0, 0)$.

\end{itemize}
Let ${\cal X}$ be the surface obtained by blowing up ${\cal X}_1$ at these $3$ points (see Figure~\ref{figure:surface} and Table~\ref{table:curves}).

\begin{table}
	\begin{tabular}{lll}
		Curve & Definition & Self-intersection \\ \hline
		$E_1$ & Strict transform of $\{ z = 0 \}$ & $- 1$ \\
		$E_2$ & Exceptional curve corresponding to (10) & $- 1$ \\
		$E_3$ & Exceptional curve corresponding to (11) & $- 1$ \\
		$E_4$ & Exceptional curve corresponding to (12) & $- 1$ \\ \hline
		$D_1$ & Strict transform of $\{ w = \infty \}$ & $- 4$ \\
		$D_2$ & Exceptional curve corresponding to (8) & $- 4$ \\
		$D_3$ & Strict transform of $\{ z = \infty \}$ & $- 4$ \\
		$D_4$ & Strict transform of $\{ w = 0 \}$ & $- 4$ \\ \hline
		$C_1$ & Exceptional curve corresponding to (1) & $- 1$ \\
		$C_2$ & Exceptional curve corresponding to (5) & $- 1$ \\
		$C_3$ & Exceptional curve corresponding to (6) & $- 1$ \\
		$C_4$ & Exceptional curve corresponding to (9) & $- 1$ \\
		$C_5$ & Exceptional curve corresponding to (7) & $- 1$ \\
		$C_6$ & Exceptional curve corresponding to (4) & $- 1$ \\
		$C_7$ & Exceptional curve corresponding to (3) & $- 1$ \\
		$C_8$ & Exceptional curve corresponding to (2) & $- 1$ \\ \hline
	\end{tabular}
	\caption{
		Curves on the surface ${\cal X}$.
		The numbers in the definition correspond to those in Figures~\ref{figure:blowup} and \ref{figure:surface}.
	}
	\label{table:curves}
\end{table}


Let us calculate degree growth for $\hat\varphi$ and $\hat\chi$.
We take
\[
	[E_1], [E_2], [E_3], [E_4], [D_1], [D_2], [D_3], [D_4], [C_2], [C_3], [C_4], [C_5], [C_6], [C_8]
\]
for a basis of $\Pic {\cal X}$, where $[-]$ denotes a class in $\Pic {\cal X}$.
The choice is made because the motion of $E_j$ and $D_j$ is cyclic, and the motion of $C_1$ and $C_7$ is complicated.
The intersection matrix on $\Pic {\cal X}$ in this basis is
\[
	A_{\cal X} := \begin{pmatrix}
		-1 & 0 & 0 & 0 & 1 & 0 & 0 & 1 & 0 & 0 & 0 & 0 & 0 & 0 \\
		 0 & -1 & 0 & 0 & 1 & 1 & 0 & 0 & 0 & 0 & 0 & 0 & 0 & 0 \\
		 0 & 0 & -1 & 0 & 0 & 1 & 1 & 0 & 0 & 0 & 0 & 0 & 0 & 0 \\
		 0 & 0 & 0 & -1 & 0 & 0 & 1 & 1 & 0 & 0 & 0 & 0 & 0 & 0 \\
		 1 & 1 & 0 & 0 & -4 & 0 & 0 & 0 & 1 & 1 & 0 & 0 & 0 & 0 \\
		 0 & 1 & 1 & 0 & 0 & -4 & 0 & 0 & 0 & 0 & 1 & 0 & 0 & 0 \\
		 0 & 0 & 1 & 1 & 0 & 0 & -4 & 0 & 0 & 0 & 0 & 1 & 0 & 0 \\
		 1 & 0 & 0 & 1 & 0 & 0 & 0 & -4 & 0 & 0 & 0 & 0 & 1 & 1 \\
		 0 & 0 & 0 & 0 & 1 & 0 & 0 & 0 & -1 & 0 & 0 & 0 & 0 & 0 \\
		 0 & 0 & 0 & 0 & 1 & 0 & 0 & 0 & 0 & -1 & 0 & 0 & 0 & 0 \\
		 0 & 0 & 0 & 0 & 0 & 1 & 0 & 0 & 0 & 0 & -1 & 0 & 0 & 0 \\
		 0 & 0 & 0 & 0 & 0 & 0 & 1 & 0 & 0 & 0 & 0 & -1 & 0 & 0 \\
		 0 & 0 & 0 & 0 & 0 & 0 & 0 & 1 & 0 & 0 & 0 & 0 & -1 & 0 \\
		 0 & 0 & 0 & 0 & 0 & 0 & 0 & 1 & 0 & 0 & 0 & 0 & 0 & -1
	\end{pmatrix}.
\]
Using this basis, we have
\begin{align*}
	H_z &= [E_2] + 2 [E_3] + [E_4] + [D_2] + [D_3] + [C_4] + [C_5], \\
	H_w &= 2 [E_2] + [E_3] + [D_1] + [D_2] + [C_2] + [C_3] + [C_4], \\
	[C_1] &= H_z - [E_1], \\
	[C_7] &= H_w - [E_4] - [D_4] - [C_6] - [C_8],
\end{align*}
where $H_z$ and $H_w$ are the classes $\{ z = (\text{const}) \}$ and $\{ w = (\text{const}) \}$, respectively.

\subsection{Degree growth}

Since the maps $\hat{\varphi}$ and $\hat{\chi}$ leave $K$ invariant, the degree growth of these maps is either quadratic or bounded.
In this subsection, we show that the degree growth is quadratic and that the maps $\hat{\varphi}$ and $\hat{\chi}$ are independent by calculating the leading coefficients for the maps and their composition.

\begin{propn}
    The degree growth of the map $\hat{\varphi}$ is quadratic, and its leading coefficient is $\frac{2}{3}$.
\end{propn}

\begin{proof}
Calculating the motion of each curve, we have the action of $\hat{\varphi}$ on $\Pic {\cal X}$ as follows:
\begin{itemize}
	\item
	$[E_1] \mapsto [E_2] \mapsto [E_3] \mapsto [E_4] \mapsto [E_1]$,

	\item
	$[D_1] \mapsto [D_2] \mapsto [D_3] \mapsto [D_4] \mapsto [D_1]$,

	\item
	$[C_2] \mapsto [C_4] \mapsto [C_5] \mapsto [C_8] \mapsto [C_3]$,

	\item
	$[C_3] \mapsto 2 H_z + H_w - ([D_2] + [C_4] + [E_2] + [E_3]) - [C_1] - [C_2] - [C_6] - [C_7]$,

	\item
	$[C_6] \mapsto H_z - [C_7]$.

\end{itemize}
Therefore, the matrix representation of the action on $\operatorname{Pic}({\cal X})$ for $\hat{\varphi}$ with respect to the basis is
\[
	M_{\hat{\varphi}} := \begin{pmatrix}
		0 & 0 & 0 & 1 &   &   &   &   & 0 & 1 & 0 & 0 & 0 & 0 \\
		1 & 0 & 0 & 0 &   &   &   &   & 0 & 0 & 0 & 0 & -1 & 0 \\
		0 & 1 & 0 & 0 &   &   &   &   & 0 & 1 & 0 & 0 & 1 & 0 \\
		0 & 0 & 1 & 0 &   &   &   &   & 0 & 2 & 0 & 0 & 2 & 0 \\
		  &   &   &   & 0 & 0 & 0 & 1 & 0 & 0 & 0 & 0 & -1 & 0 \\
		  &   &   &   & 1 & 0 & 0 & 0 & 0 & 0 & 0 & 0 & 0 & 0 \\
		  &   &   &   & 0 & 1 & 0 & 0 & 0 & 1 & 0 & 0 & 1 & 0 \\
		  &   &   &   & 0 & 0 & 1 & 0 & 0 & 1 & 0 & 0 & 1 & 0 \\
		  &   &   &   &   &   &   &   & 0 & -1 & 0 & 0 & -1 & 0 \\
		  &   &   &   &   &   &   &   & 0 & 0 & 0 & 0 & -1 & 1 \\
		  &   &   &   &   &   &   &   & 1 & 0 & 0 & 0 & 0 & 0 \\
		  &   &   &   &   &   &   &   & 0 & 1 & 1 & 0 & 1 & 0 \\
		  &   &   &   &   &   &   &   & 0 & 0 & 0 & 0 & 1 & 0 \\
		  &   &   &   &   &   &   &   & 0 & 1 & 0 & 1 & 1 & 0
	\end{pmatrix},
\]
where the entries of the empty blocks are all zero.

To compute the degree growth and the leading coefficient for $\hat{\varphi}$, we need to calculate
\[
	\lim_{n \to \infty} \frac{1}{n^2} M^2_{\hat{\varphi}}.
\]
Let
\[
	H_{\cal X} := [E_1] + [E_2] + [D_1] + [C_1] + [C_2] + [C_3],
\]
which is the total transform of the class of lines in $\mathbb{P}^2$ (see \cite[Appendix~A]{hkm24}), and let
\[
	h_{\cal X} := (0, 2, 2, 1, 1, 1, 1, 0, 1, 1, 1, 1, 0, 0)^T \in \mathbb{Z}^{14},
\]
which corresponds to $H_{\cal X}$ under the basis.
Then,
\[
	\deg \left( \hat{\varphi}^n \right) = \left\langle h_{\cal X}, M^n_{\hat{\varphi}} h_{\cal X} \right\rangle_{A_{\cal X}}
\]
where $\langle -, - \rangle_{A_{\cal X}}$ is the bilinear form defined by $A_{\cal X}$:
\[
	\langle u_1, u_2 \rangle_{A_{\cal X}} = u^T_1 A_{\cal X} u_2
\]
for $u_1, u_2 \in \mathbb{Z}^{14}$.
Therefore, we have
\[
	\lim_{n \to \infty} \frac{1}{n^2} \deg \left( \hat{\varphi}^n \right)
	= \left\langle h_{\cal X}, \lim_{n \to \infty} \left( \frac{1}{n^2} M^n_{\hat{\varphi}} \right) h_{\cal X} \right\rangle_{A_{\cal X}}.
\]
However, because of the large size of $M^n_{\hat{\varphi}}$, it is not easy to directly calculate the Jordan basis for $M^n_{\hat{\varphi}}$ 
with computer algebra. 
Here, we will use the general theory from \cite{mase17}.

A direct calculation shows that the characteristic polynomial of $M_{\hat{\varphi}}$ is
\[
	(\lambda - 1)^5 (\lambda + 1)^3 (\lambda^2 + 1) (\lambda^2 + \lambda + 1)
\]
and
\[
	(M_{\hat{\varphi}} - I_{14}) (M_{\hat{\varphi}} + I_{14}) (M^2_{\hat{\varphi}} + I_{14}) (M^2_{\hat{\varphi}} + M_{\hat{\varphi}} + I_{14}) \ne O,
\]
where $I_{14}$ is the identity matrix of size $14$.
By \cite[Proposition~3.2]{mase17}, the degree growth for the map is quadratic and the minimal polynomial of $M_{\hat{\varphi}}$ is
\[
	(\lambda - 1)^3 (\lambda + 1) (\lambda^2 + 1) (\lambda^2 + \lambda + 1).
\]
Therefore, the matrix
\[
	(M_{\hat{\varphi}} + I_{14}) (M^2_{\hat{\varphi}} + I_{14}) (M^2_{\hat{\varphi}} + M_{\hat{\varphi}} + I_{14})
	= \begin{pmatrix}
		3 & 3 & 3 & 3 &   &   &   &   & 1 & 25 & 4 & 9 & 10 & 16 \\
		3 & 3 & 3 & 3 &   &   &   &   & 0 & 16 & 1 & 4 & -5 & 9 \\
		3 & 3 & 3 & 3 &   &   &   &   & 1 & 17 & 2 & 5 & 2 & 10 \\
		3 & 3 & 3 & 3 &   &   &   &   & 2 & 26 & 5 & 10 & 17 & 17 \\
		  &   &   &   & 3 & 3 & 3 & 3 & 0 & 10 & 1 & 3 & -2 & 6 \\
		  &   &   &   & 3 & 3 & 3 & 3 & 0 & 6 & 0 & 1 & -3 & 3 \\
		  &   &   &   & 3 & 3 & 3 & 3 & 1 & 11 & 2 & 4 & 5 & 7 \\
		  &   &   &   & 3 & 3 & 3 & 3 & 1 & 15 & 3 & 6 & 12 & 10 \\
		  &   &   &   &   &   &   &   & 0 & -8 & -2 & -4 & -8 & -6 \\
		  &   &   &   &   &   &   &   & 2 & 10 & 4 & 6 & -2 & 8 \\
		  &   &   &   &   &   &   &   & 2 & -6 & 0 & -2 & -6 & -4 \\
		  &   &   &   &   &   &   &   & 4 & 4 & 4 & 4 & 4 & 4 \\
		  &   &   &   &   &   &   &   & 0 & 0 & 0 & 0 & 12 & 0 \\
		  &   &   &   &   &   &   &   & 4 & 12 & 6 & 8 & 12 & 10
	\end{pmatrix}
\]
acts on $\mathbb{C}^{14}$ as a projection to the generalized eigenspace for $1$.
A direct calculation shows that the $9$th column of the matrix
\[
	(M_{\hat{\varphi}} - I_{14})^2 (M_{\hat{\varphi}} + I_{14}) (M^2_{\hat{\varphi}} + I_{14}) (M^2_{\hat{\varphi}} + M_{\hat{\varphi}} + I_{14})
\]
is non-zero.
Let $v_3 \in \mathbb{Z}^{14}$be the $9$th column of $(M_{\hat{\varphi}} + I_{14}) (M^2_{\hat{\varphi}} + I_{14}) (M^2_{\hat{\varphi}} + M_{\hat{\varphi}} + I_{14})$, i.e.,
\[
	v_3 = (1, 0, 1, 2, 0, 0, 1, 1, 0, 2, 2, 4, 0, 4)^T,
\]
and let
\[
	v_2 = (M_{\hat{\varphi}} - I_{14}) v_3, \quad
	v_1 = (M_{\hat{\varphi}} - I_{14}) v_2.
\]
Then, $v_1 \ne 0$ and $(v_1, v_2, v_3)$ corresponds to the Jordan block of size $3$ for $M_{\hat{\varphi}}$.

By \cite[Lemma~3.7]{mase17}, it holds that for $u \in \mathbb{C}^{14}$,
\[
	\lim_{n \to \infty} \frac{1}{n^2} M^n_{\hat{\varphi}} u = \frac{\langle u, v_1 \rangle_{A_{\cal X}}}{2 \langle v_3, v_1 \rangle_{A_{\cal X}}} v_1.
\]
Therefore, we have
\begin{align*}
	\lim_{n \to \infty} \frac{1}{n^2} \deg \left( \hat{\varphi}^n \right)
	&= \left\langle h_{\cal X}, \frac{\langle h_{\cal X}, v_1 \rangle_{A_{\cal X}}}{2 \langle v_3, v_1 \rangle_{A_{\cal X}}} v_1 \right\rangle_{A_{\cal X}} \\
	&= \frac{\langle h_{\cal X}, v_1 \rangle^2_{A_{\cal X}}}{2 \langle v_3, v_1 \rangle_{A_{\cal X}}} \\
	&= \frac{2}{3}.
\end{align*}
\end{proof}

\begin{propn}
    The degree growth of the map $\hat{\chi}$ is quadratic, and its leading coefficient is $\frac{14}{3}$.
\end{propn}

\begin{proof}
The action of $\hat{\chi}$ on $\Pic {\cal X}$ is
\begin{itemize}
	\item
	$[E_1] \mapsto [E_2] \mapsto [E_3] \mapsto [E_4] \mapsto [E_1]$,

	\item
	$[D_1] \mapsto [D_2] \mapsto [D_3] \mapsto [D_4] \mapsto [D_1]$,

	\item
	$[C_2] \mapsto 2 H_z + H_w - ([D_2] + [C_4] + [E_2] + [E_3]) - [C_1] - [C_3] - [C_6] - [C_8]$,
	
	\item
	$[C_3] \mapsto H_z + H_w - ([D_2] + [C_4] + [E_2] + [E_3]) - [C_1] - [C_6]$,

	\item
	$[C_4] \mapsto H_z + H_w - [C_1] - [C_3] - [C_6]$,

	\item
	$[C_5] \mapsto [C_7]$,

	\item
	$[C_6] \mapsto [C_2]$,

	\item
	$[C_8] \mapsto H_z - [C_6]$

\end{itemize}
and the matrix representation is
\[
	M_{\hat{\chi}} := \begin{pmatrix}
		0 & 0 & 0 & 1 &   &   &   &   & 1 & 1 & 1 & 0 & 0 & 0 \\
		1 & 0 & 0 & 0 &   &   &   &   & 2 & 1 & 2 & 2 & 0 & 1 \\
		0 & 1 & 0 & 0 &   &   &   &   & 2 & 0 & 1 & 1 & 0 & 2 \\
		0 & 0 & 1 & 0 &   &   &   &   & 1 & 0 & 0 & -1 & 0 & 1 \\
		  &   &   &   & 0 & 0 & 0 & 1 & 1 & 1 & 1 & 1 & 0 & 0 \\
		  &   &   &   & 1 & 0 & 0 & 0 & 1 & 0 & 1 & 1 & 0 & 1 \\
		  &   &   &   & 0 & 1 & 0 & 0 & 1 & 0 & 0 & 0 & 0 & 1 \\
		  &   &   &   & 0 & 0 & 1 & 0 & 0 & 0 & 0 & -1 & 0 & 0 \\
		  &   &   &   &   &   &   &   & 1 & 1 & 1 & 1 & 1 & 0 \\
		  &   &   &   &   &   &   &   & 0 & 1 & 0 & 1 & 0 & 0 \\
		  &   &   &   &   &   &   &   & 1 & 0 & 1 & 1 & 0 & 1 \\
		  &   &   &   &   &   &   &   & 1 & 0 & 0 & 0 & 0 & 1 \\
		  &   &   &   &   &   &   &   & -1 & -1 & -1 & -1 & 0 & -1 \\
		  &   &   &   &   &   &   &   & -1 & 0 & 0 & -1 & 0 & 0
	\end{pmatrix}.
\]
A direct calculation shows that $M_{\hat{\varphi}}$ and $M_{\hat{\chi}}$ have the same characteristic polynomial.
One can show by calculation that
\[
	(M_{\hat{\chi}} + I_{14}) (M^2_{\hat{\chi}} + I_{14}) (M^2_{\hat{\chi}} + M_{\hat{\chi}} + I_{14})
	= \begin{pmatrix}
		3 & 3 & 3 & 3 &   &   &   &   & 65 & 49 & 58 & 53 & 34 & 50 \\
		3 & 3 & 3 & 3 &   &   &   &   & 104 & 80 & 95 & 88 & 59 & 83 \\
		3 & 3 & 3 & 3 &   &   &   &   & 87 & 63 & 84 & 71 & 48 & 72 \\
		3 & 3 & 3 & 3 &   &   &   &   & 48 & 32 & 47 & 36 & 23 & 39 \\
		  &   &   &   & 3 & 3 & 3 & 3 & 51 & 41 & 44 & 44 & 29 & 39 \\
		  &   &   &   & 3 & 3 & 3 & 3 & 53 & 39 & 51 & 44 & 30 & 44 \\
		  &   &   &   & 3 & 3 & 3 & 3 & 34 & 24 & 33 & 27 & 18 & 28 \\
		  &   &   &   & 3 & 3 & 3 & 3 & 14 & 8 & 14 & 9 & 5 & 11 \\
		  &   &   &   &   &   &   &   & 28 & 20 & 20 & 24 & 20 & 16 \\
		  &   &   &   &   &   &   &   & 10 & 18 & 6 & 14 & 6 & 10 \\
		  &   &   &   &   &   &   &   & 24 & 16 & 28 & 20 & 16 & 24 \\
		  &   &   &   &   &   &   &   & 4 & 4 & 4 & 4 & 4 & 4 \\
		  &   &   &   &   &   &   &   & -30 & -30 & -30 & -30 & -18 & -30 \\
		  &   &   &   &   &   &   &   & -24 & -16 & -16 & -20 & -16 & -12
	\end{pmatrix}
\]
and the $9$th column of the matrix
\[
	(M_{\hat{\chi}} - I_{14})^2 (M_{\hat{\chi}} + I_{14}) (M^2_{\hat{\chi}} + I_{14}) (M^2_{\hat{\chi}} + M_{\hat{\chi}} + I_{14})
\]
is non-zero.
Let
\[
	v_3 = (65, 104, 87, 48, 51, 53, 34, 14, 28, 10, 24, 4, -30, -24)^T
\]
and let
\[
	v_2 = (M_{\hat{\chi}} - I_{14}) v_3, \quad
	v_1 = (M_{\hat{\chi}} - I_{14}) v_2.
\]
Then, we have
\begin{align*}
	\lim_{n \to \infty} \frac{1}{n^2} \deg \left( \hat{\chi}^n \right)
	&= \left\langle h_{\cal X}, \frac{\langle h_{\cal X}, v_1 \rangle_{A_{\cal X}}}{2 \langle v_3, v_1 \rangle_{A_{\cal X}}} v_1 \right\rangle_{A_{\cal X}} \\
	&= \frac{\langle h_{\cal X}, v_1 \rangle^2_{A_{\cal X}}}{2 \langle v_3, v_1 \rangle_{A_{\cal X}}} \\
	&= \frac{14}{3}.
\end{align*}
\end{proof}

\begin{propn}
    The degree growth of the composition map $\hat{\varphi} \cdot \hat{\chi}$ is quadratic, and its leading coefficient is $4$.
\end{propn}

\begin{proof}
Let
\[
	M := M_{\hat{\varphi}} M_{\hat{\chi}}
	= \begin{pmatrix}
		0 & 0 & 1 & 0 &   &   &   &   & 1 & 1 & 0 & 0 & 0 & 1 \\
		0 & 0 & 0 & 1 &   &   &   &   & 2 & 2 & 2 & 1 & 0 & 1 \\
		1 & 0 & 0 & 0 &   &   &   &   & 1 & 1 & 1 & 2 & 0 & 0 \\
		0 & 1 & 0 & 0 &   &   &   &   & 0 & 0 & -1 & 1 & 0 & 0 \\
		  &   &   &   & 0 & 0 & 1 & 0 & 1 & 1 & 1 & 0 & 0 & 1 \\
		  &   &   &   & 0 & 0 & 0 & 1 & 1 & 1 & 1 & 1 & 0 & 0 \\
		  &   &   &   & 1 & 0 & 0 & 0 & 0 & 0 & 0 & 1 & 0 & 0 \\
		  &   &   &   & 0 & 1 & 0 & 0 & 0 & 0 & -1 & 0 & 0 & 0 \\
		  &   &   &   &   &   &   &   & 1 & 0 & 1 & 0 & 0 & 1 \\
		  &   &   &   &   &   &   &   & 0 & 1 & 1 & 0 & 0 & 1 \\
		  &   &   &   &   &   &   &   & 1 & 1 & 1 & 1 & 1 & 0 \\
		  &   &   &   &   &   &   &   & 0 & 0 & 0 & 1 & 0 & 0 \\
		  &   &   &   &   &   &   &   & -1 & -1 & -1 & -1 & 0 & -1 \\
		  &   &   &   &   &   &   &   & 0 & 0 & -1 & 0 & 0 & 0
	\end{pmatrix},
\]
which is the matrix representation of the action on $\Pic {\cal X}$ of $\hat{\varphi} \cdot \hat{\chi}$.
A direct calculation shows that the characteristic polynomial of this matrix is
\[
	(\lambda - 1)^9 (\lambda + 1)^5.
\]
A direct calculation shows that the $9$th column of the matrix
\[
	(M - I_{14})^2 (M + I_{14})
\]
is non-zero.
Let
\[
	v_3 = (1, 2, 1, 0, 1, 1, 0, 0, 2, 0, 1, 0, -1, 0)^T,
\]
which is the $9$th column of $(M + I_{14})$, and let
\[
	v_2 = (M - I_{14}) v_3, \quad
	v_1 = (M - I_{14}) v_2.
\]
Then, we have
\begin{align*}
	\lim_{n \to \infty} \frac{1}{n^2} \deg \left( \left( \hat{\varphi} \cdot \hat{\chi} \right)^n \right)
	&= \left\langle h_{\cal X}, \frac{\langle h_{\cal X}, v_1 \rangle_{A_{\cal X}}}{2 \langle v_3, v_1 \rangle_{A_{\cal X}}} v_1 \right\rangle_{A_{\cal X}} \\
	&= \frac{\langle h_{\cal X}, v_1 \rangle^2_{A_{\cal X}}}{2 \langle v_3, v_1 \rangle_{A_{\cal X}}} \\
	&= 4.
\end{align*}
\end{proof}

\begin{thm}
    The maps $\hat{\varphi}$ and $\hat{\chi}$ are independent.
\end{thm}

\begin{proof}
The proof follows the same argument as in \cite[Appendix~A]{hkm24}.
Define the canonical height $\hat{h}$ and associated bilinear pairing $<-, ->$ on the Mordell-Weil group in the usual way, 
and let $P_1$ and $P_2$ be the translations on ${\cal X}$ (considered as an elliptic curve over $\mathbb{C}(\ka)$) corresponding to $\hat{\varphi}$ and $\hat{\chi}$, respectively.
Then we have
\begin{align*}
	\hat{h}(P_1) &= \frac{1}{3} \lim_{n \to \infty} \frac{1}{n^2} \deg \left( \hat{\varphi}^n \right) = \frac{2}{9}, \\
	\hat{h}(P_2) &= \frac{1}{3} \lim_{n \to \infty} \frac{1}{n^2} \deg \left( \hat{\chi}^n \right) = \frac{14}{9}, \\
	\hat{h}(P_1 + P_2) &= \frac{1}{3} \lim_{n \to \infty} \frac{1}{n^2} \deg \left( \hat{\varphi}^n \hat{\chi}^n \right) = \frac{4}{3}
\end{align*}
and
$$ 
	<P_1, P_1> = \frac{4}{9}, \quad 
	<P_2, P_2> = \frac{28}{9}, \quad
	<P_1, P_2> = - \frac{4}{9}.
$$ 
Therefore, the associated Gram matrix is
\beq \label{gram}
	\begin{pmatrix}
		<P_1, P_1> & <P_1, P_2> \\
		<P_2, P_1> & <P_2, P_2>
	\end{pmatrix}
	= \begin{pmatrix}
		4 / 9 & - 4 / 9 \\
		- 4 / 9 & 28 / 9
	\end{pmatrix}.
\eeq
Since this matrix is non-singular, the maps $\hat{\varphi}$ and $\hat{\chi}$ are independent.
\end{proof}

\begin{remark} Observe that the ratio of coefficients in the quadratic form associated with \eqref{gram} agrees with the result of Theorem~\ref{degrowth}. Indeed, by a more involved calculation, one can recover the exact formula for the degree of $\hat{\chi}^m\hat{\varphi}^n$ from the degrees of the cluster variables on $\Z^2$.  
\end{remark}

\subsection{QRT map}

Since it does not have bidegree $(2, 2)$ with respect to $(z, w)$, the invariant $K$ does not generate a QRT map on this chart.
However, if we find a chart on which the invariant has bidegree $(2, 2)$, we obtain a QRT map.
In this subsection, we construct such a chart with the help of the space of initial conditions.

The invariant is not of bidegree $(2, 2)$ because ${\cal X}_1$, which is obtained by blowing up all the indeterminacies of the pencil, is not minimal as a rational elliptic surface.
Therefore, we need to find a chart on which $\{ z = 0 \}$ is a point.

Let ${\cal X}_{-1}$ be the surface obtained by blowing up $\mathbb{P}^1 \times \mathbb{P}^1$ at points (1) and (8).
Then, ${\cal X}_{-1}$ is isomorphic to the del Pezzo surface of degree $6$, which is unique up to isomorphism.
The surface ${\cal X}_{-1}$ has $6$ exceptional curves of first kind:
\begin{itemize}
	\item
	the exceptional curves corresponding to (1) and (6), respectively,

	\item
	the strict transforms of the curves $\{ z = 0 \}$, $\{ z = \infty \}$, $\{ w = - 1 \}$ and $\{ w = \infty \}$, respectively.

\end{itemize}
The configuration of these $6$ curves forms the Dynkin diagram of type $A^{(1)}_5$, and the group of all the automorphisms of ${\cal X}_{-1}$ coincides with the group of automorphisms of the Dynkin diagram.
Each automorphism of ${\cal X}_{-1}$ is written by a linear transformation in $z$ and $w + 1$ w.r.t.\ multiplication of order $6$.

The matrix
$\begin{pmatrix}
	1 & 0 \\
	1 & - 1
\end{pmatrix}$
has order $2$ and thus corresponds to an automorphism (involution) of ${\cal X}_{-1}$:
\[
	\begin{pmatrix}
		z \\ w 
	\end{pmatrix}
	\mapsto \begin{pmatrix}
		z \\
		\frac{z}{w + 1}-1
	\end{pmatrix}.
\]
The motion of each exceptional curve under this automorphism is the following:
\begin{itemize}
	\item
	$\{ z = 0 \}$ $\leftrightarrow$ exceptional curve of (1),

	\item
	$\{ w = - 1 \}$ $\leftrightarrow$ $\{ w = \infty \}$,

	\item
	$\{ z = \infty \}$ $\leftrightarrow$ exceptional curve of (8).

\end{itemize}
Therefore, this automorphism makes the curve $\{ z = 0 \}$ invisible on $\mathbb{P}^1 \times \mathbb{P}^1$.

Based on this automorphism, we define a new chart $(\widetilde{z}, \widetilde{w})$ as
\[
	\widetilde{z} = z, \quad
	\widetilde{w} = \frac{z}{w + 1}.
\]
Note that we omit  a shift by $- 1$ in $\widetilde{w}$ because it does not change the bidegree of the invariant.
As expected, on the chart $(\widetilde{z}, \widetilde{w})$, the invariant $K$ has bidegree $(2, 2)$:
\[
	K = \frac{\left( - \alpha - \beta -2 \right) \widetilde{w}^2 + \left( - \widetilde{z}^2 - \alpha - \beta + 2 \widetilde{z} + 1 \right) \widetilde{w} - \left( \widetilde{z} + 1 \right) \left( \alpha \beta + \widetilde{z} \right)}{\left( - \widetilde{z} + \widetilde{w} \right) \widetilde{w}}.
\]
Since it has bidegree $(2, 2)$, the invariant generates a QRT map $\varphi_{\mathrm{QRT}}$ on this chart. To see this, note that, on each biquadratic curve in the pencil defined by $K=\,$const, we have two natural involutions, 
which 
 in the terminology of Duistermaat \cite{Duistermaat2010DiscreteIS} are called 
 the vertical switch, that is  
$$ 
\iota_v: \quad 
\left(\begin{array}{c} \widetilde{z} \\ 
\widetilde{w} \end{array} \right) \mapsto 
\left(\begin{array}{c} \widetilde{z} \\ 
\widetilde{w}^* \end{array} \right),  
$$
where 
$$ 
\widetilde{w}^* 
 = \frac{ (\widetilde{z}-\widetilde{w})(\widetilde{z}+\al\be)}  {\widetilde{w}\widetilde{z}+\widetilde{z}+\al\be+(\al+\be-1)\widetilde{w}}  , 
$$
and the horizontal switch, given by 
$$ 
\iota_h: \quad 
\left(\begin{array}{c} \widetilde{z} \\ 
\widetilde{w} \end{array} \right) \mapsto 
\left(\begin{array}{c} \widetilde{z}^* \\ 
\widetilde{w} \end{array} \right),  
$$
where 
$$ \widetilde{z}^* 
 = \frac{\widetilde{w}\widetilde{z}+(\al+\be) \widetilde{w} +\al\be}{(\widetilde{z}-\widetilde{w})}. 
$$
Then we can define the composition 
$$ 
\varphi_{\mathrm{QRT}} = \iota_h\cdot \iota_v, 
$$ 
namely 
\[
	\varphi_{\mathrm{QRT}} \colon\, 
	\begin{pmatrix}
		\widetilde{z} \\
		\widetilde{w}
	\end{pmatrix}
	\mapsto 
    \left(\begin{array}{c} \widetilde{z}^* \\ 
\widetilde{w}^* \end{array} \right) = 
	\begin{pmatrix}
    \dfrac{\widetilde{w}^*\widetilde{z}+(\al+\be) \widetilde{w}^* +\al\be}{(\widetilde{z}-\widetilde{w}^*)}
	     \\[1em] 
         \dfrac{ (\widetilde{z}-\widetilde{w})(\widetilde{z}+\al\be)}  {\widetilde{w}\widetilde{z}+\widetilde{z}+\al\be+(\al+\be-1)\widetilde{w}}
	\end{pmatrix}
    .
\]
The following coincidence is remarkable.

\begin{propn}
    If rewritten back in the chart $(z, w)$, the QRT map $\varphi_{\mathrm{QRT}}$ coincides with the deformed D$_4$ map $\hat{\varphi}$.
\end{propn}

\begin{proof}
    By direct calculation.
\end{proof}


\section{Conclusions}

Our study, initiated in \cite{hk} and developed further in \cite{hkm24}, has revealed numerous instances where simple deformations of mutations in finite Dynkin type cluster algebras produce parameter families of maps that are integrable in the Liouville sense. While these deformations destroy Zamolodchikov periodicity and the Laurent property in the original cluster variables, the resulting maps appear to admit a lift to an extended space of cluster variables, in which the Laurent property is restored. Aside from many particular examples, we have so far succeeded in carrying out this programme more systematically for all type $\rA$ algebras of even rank \cite{grab}, and more recently one of us has found a similar construction for even rank type $\rD$ algebras \cite{kim2025}. 

The focus of this paper has been on the complex geometry of the integrable symplectic map defined by deformed mutations in the cluster algebra of type $\rD_4$.  In the standard setting of Hamiltonian mechanics on a real symplectic manifold, Liouville integrability entails that the compact connected level sets of the first integrals are diffeomorphic to tori, while in the more refined  context of algebraic integrability over $\C$, Liouville tori are promoted to abelian varieties. For the case at hand, with one degree of freedom, we have seen that generic level sets of the first integral for the deformed $\rD_4$ symplectic map $\hat\varphi$ are elliptic curves, and the whole family of these level sets is a rational elliptic surface. Remarkably, the extended cluster structure for the lift of the deformed map $\hat\varphi$ appears to carry all the information about the arithmetic of the Mordell-Weil group: this has rank 2, with two independent generators of translations on the fibres of the surface that are both decomposed into cluster mutations. In \cite{hkm24} we found an analogous structure in the case of the rational elliptic surface associated  with a deformed cluster map in type $\rA_3$. 

There are still many open questions to consider regarding these deformed cluster maps. In particular, we previously found another integrable deformation of the $\rD_4$ cluster map, corresponding to case (1) in Theorem~\ref{defd4}, which has a different extended cluster structure from the map $\hat\varphi$ considered here. Nevertheless, the mutable part of the quiver for the two maps (cases (1) and (2) in the theorem) turns out to be the same, so the relation between them remains to be understood. Moreover, the same mutable quiver is mutation equivalent to the one used by Okubo to derive the $q$-Painlev\'e VI equation from Y-seed mutations \cite{okubo}, so the connection of these deformed cluster maps with  the corresponding Y-systems is worthy of further analysis. Finally, the geometry of deformed cluster maps for Dynkin type algebras of higher rank should be even more interesting, with the level sets of integrals yielding families of higher-dimensional abelian varieties, with torsion points appearing upon reduction to the 
original (undeformed) setting of Zamolodchikov periodicity.

\section*{Acknowledgments}

This research was supported by
a Grant-in-Aid for Scientific Research of Japan Society for the Promotion of Science, JSPS KAKENHI Grant Number 23K12996 and 24KF0208. 
We used Bernhard Keller's quiver mutation application to produce Figure 1  - see the webpage 

{\tt https://webusers.imj-prg.fr/\~{}bernhard.keller/quivermutation/} 

\noindent 
for details. 
ANWH is grateful to BIMSA for supporting his stay there during the RTISART conference in July 2024, and to the Graduate School of Mathematical Sciences, University of Tokyo for the invitation to speak at the FoPM Symposium in February 2025, where this paper was first conceived. He would also like to thank Tetsushi Ito (Kyoto) for supporting his subsequent visit to Japan in June, where the Number Theory \& Integrable Systems workshop in Kobe provided a further opportunity for us to work on the paper together. The authors would like to thank an anonymous referee for helpful suggestions which improved the clarity of the paper.



\begin{thebibliography}{99}

\bibitem{Duistermaat2010DiscreteIS} 
J.J. Duistermaat,
Discrete Integrable Systems: QRT Maps and Elliptic Surfaces, Springer Monographs in Mathematics, 2010. 

\bibitem{fz2} 
S.~Fomin and A.~Zelevinsky,  
Cluster algebras. II. Finite type classification, Invent. Math.  154 (2003) 63--121.

\bibitem{fzYsys}
S.~Fomin and A.~Zelevinsky, Y-systems and generalized associahedra, Ann. Math. 158
(2003) 977–1018

\bibitem{fz4} 
S.~Fomin and A.~Zelevinsky,  
Cluster algebras: IV. Coefficients, Compos. Math. 143 (2007) 112--64.


\bibitem{grab} J.E. Grabowski, A.N.W. Hone and W. Kim, 
Deformed cluster maps of type $\rA_{2N}$, preprint (2024); 
{\tt arXiv:2402.18310} 

\bibitem{hk} A.N.W. Hone and T.E. Kouloukas,
Deformations of cluster mutations and invariant presymplectic forms, 
J. Alg. Comb. 57 (2023) 763--791.



\bibitem{hkm24}
	A.N.W. Hone, W. Kim, T. Mase,
	New cluster algebras from old: integrability beyond Zamolodchikov periodicity,
	\textit{Journal of Physics A: Mathematical and Theoretical}
	57
	(2024) 
	415201.

\bibitem{kim2025}
W.Kim,
Integrable deformations of cluster maps of type $\rD_{2N}$, preprint (2025);
{\tt arXiv:2506.06182}


\bibitem{kun} A. Kuniba, T. Nakanishi and J.  Suzuki, 
T-systems and Y-systems in integrable systems, J. Phys. A: Math. Theor. 44 (2011) 103001.

	\bibitem{mase17}
	T. Mase,
	Studies on spaces of initial conditions for nonautonomous mappings of the plane,
	\textit{Journal of Integrable Systems}
	3
	(2018):
	xyy010.

\bibitem{MGfriezes}
S. Morier-Genoud, 
Coxeter’s frieze patterns at the crossroads of algebra,
geometry and combinatorics, 
Bull. London Math. Soc. 47 (2015) 895--938. 

\bibitem{os} K. Oguiso and T. Shioda,  
The Mordell-Weil Lattice of a Rational Elliptic Surface, 
Commentarii Mathematici Universitatis Sancti Pauli 40 
(1991) 83--99. 

\bibitem{okubo} 
N. Okubo, 
Bilinear equations and
q-discrete Painlevé
equations satisfied by variables and coefficients in
cluster algebras, 
J. Phys. A: Math. Theor. 48 (2015) 355201. 

\bibitem{zam} Al.B. Zamolodchikov, 
On the thermodynamic Bethe ansatz equations for reflectionless ADE scattering theories,
Phys. Lett. B 253 (1991) 391--394.

 

\end{thebibliography}
\end{document}